\documentclass[journal,10pt]{IEEEtran}

\usepackage{cite}
\usepackage{color} 
\usepackage{bm}
\usepackage{subfigure}
\usepackage{multirow}

\ifCLASSINFOpdf
   \usepackage[pdftex]{graphicx}
\else
\fi
\usepackage{cite}
\usepackage{amsmath,amssymb,graphicx,algorithm,algorithmic}
\usepackage{makecell}

\usepackage{color}   
\usepackage{array}
\begin{document}

\title{A Cluster-Based Statistical Channel Model for Integrated Sensing and Communication Channels}

\author{Zhengyu~Zhang,~\IEEEmembership{Student Member, IEEE,}
        Ruisi~He,~\IEEEmembership{Senior Member, IEEE,}
        Bo~Ai,~\IEEEmembership{Fellow, IEEE,}
        Mi~Yang,~\IEEEmembership{Member, IEEE,}
        Yong~Niu,~\IEEEmembership{Member, IEEE,}
        Zhangdui~Zhong,~\IEEEmembership{Fellow, IEEE,}
        Yujian~Li,~\IEEEmembership{Member, IEEE,}
        Xuejian~Zhang
        and Jing~Li~\IEEEmembership{}

\thanks{


Zhengyu Zhang, Ruisi He, Bo Ai, Mi Yang, Yong Niu, Zhangdui Zhong and Xuejian Zhang are with the State Key Laboratory of Advanced Rail Autonomous Operation, Beijing Jiaotong University, Beijing 100044, China

Yujian Li and Jing Li are with the Key Laboratory of All Optical Network and Advanced Telecommunication Network of EMC, Institute of Lightwave Technology, Beijing Jiaotong University, Beijing 100044, China

              }
\thanks{
              }}

\markboth{}
{}

\maketitle

\begin{abstract}
The emerging 6G network envisions integrated sensing and communication (ISAC) as a promising solution to meet growing demand for native perception ability. To optimize and evaluate ISAC systems and techniques, it is crucial to have an accurate and realistic wireless channel model. However, some important features of ISAC channels have not been well characterized, for example, most existing ISAC channel models consider communication channels and sensing channels independently, whereas ignoring correlation under the consistent environment. Moreover, sensing channels have not been well modeled in the existing standard-level channel models. Therefore, in order to better model ISAC channel, a cluster-based statistical channel model is proposed in this paper, which is based on measurements conducted at 28 GHz. In the proposed model, a new framework based on 3GPP standard is proposed, which includes communication clusters and sensing clusters. Clustering and tracking algorithms are used to extract and analyze ISAC channel characteristics. Furthermore, some special sensing cluster structures such as shared sensing cluster, newborn sensing cluster, etc., are defined to model correlation and difference between communication and sensing channels. Finally, accuracy of the proposed model is validated based on measurements and simulations.

\end{abstract}


\IEEEpeerreviewmaketitle

\section{Introduction}

\IEEEPARstart{T}{H}e sixth generation (6G) wireless network is being developed to revolutionize communication paradigm, enabling ubiquitous sensing, connectivity, and intelligence \cite{Ref1}. To meet higher demands for end-to-end information processing of 6G, there is a growing interest in integrated sensing and communication (ISAC) techniques, which enables native perception ability to support smart homes, enhanced positioning and environmental monitoring, etc \cite{Ref2,Ref3}. Meanwhile, the development of wireless communication towards higher frequencies such as millimeter waves, terahertz waves, and visible light will result in more overlaps with traditional radar frequencies. Achieving communication and sensing in the same spectrum can avoid interference and improve spectrum utilization. Furthermore, in recent years, wireless communication and radar sensing have more similarities in system design, signal processing, and data transmission. Sharing the same software and hardware equipment to achieve communication and sensing can reduce equipment cost, size, and power consumption. Therefore, ISAC has the potential to improve spectrum and reduce cost through integration of sensing and communication functionalities in a single transmission, device, or network infrastructure \cite{Ref4}, and thus supports a lot of emerging applications in 6G.

Accurate channel models are the prerequisite for design and deployment of wireless communication systems \cite{Ref5,Ref6,Ref7}. For ISAC technologies, design, beamforming, and signal processing have been widely investigated \cite{Ref8,Ref9,Ref10}. However, propagation of communication radios and sensing echoes, i.e., ISAC channel, has not been well investigated, which is an essential foundation for ISAC system desgn and evaluation \cite{Ref11}. Generally speaking, ISAC channels integrate sensing and communication channels. For passive perception, ISAC channels are similar to the traditional communication channels. However, for active perception, ISAC channels involve radio propagation from transmitter to receiver as well as echo propagation from transmitter to scatterers and coming back to the transmitter. Therefore, traditional channel models are inappropriate for ISAC scenarios \cite{Ref12,Ref13}, and it is necessary to develop accurate ISAC channel models.

Compared to non-clustered channel models, cluster-based models are widely applied and have been adopted by many standards models such as ITU-R, 3GPP, WINNER and QuaDRiGa channel models. They separately describe inter-cluster and intra-cluster properties, making the model significantly less complex without sacrificing accuracy. Currently, channel model for ISAC is far from standardized modeling procedure, and the use of a cluster-based channel model for ISAC channels is critical. Besides, actual measurements for ISAC channel have shown that MPCs are generally distributed in clusters. Different clusters can be mapped to different scatterers in the environment, which is beneficial for modeling. Motivated by the above preference, this paper proposes using cluster-based channel model for ISAC.

\subsection{Related Work}

Modeling of ISAC channels needs to consider characteristics of sensing and communication channels. To the best of the authors' knowledge, there exist few measurements and channel modeling for ISAC scenarios. This subsection summarizes some related investigations on i) measurement campaign of ISAC, ii) independently modeling of ISAC channels, and iii) correlatively modeling of ISAC channels, as follows.

\begin{itemize}
\item[1)] 
Measurement is an effective method to characterize wireless channel \cite{Ref14,Ref15,Ref16}. In order to explore special characteristics of sensing channels, there exists some measurement campaigns for ISAC scenarios, which mainly extract channel parameters based on radar echoes. In \cite{Ref17}, an empirical statistical model for 77 GHz is presented based on measurements of onboard millimeter wave radar in underground parking lot scenarios, and analyzes multipath components characteristics using Space-Alternating Generalized Expectation-maximization (SAGE) algorithm. Ref.\cite{Ref18} conducts air-to-ground measurements in an open environment based on 5G-NR waveforms, and analyzes angular characteristics of sensing channel; In \cite{Ref19}, a radar multipath model is proposed, and amplitude and Doppler characteristics of sensing channel in indoor corner scenario are analyzed; Ref.\cite{Ref20} extracts channel parameters based on 28 GHz outdoor measurement and introduces metal scatterers to investigate impact of interference on sensing channels. However, these measurement-based ISAC channel characterizations only focus on sensing channel, which cannot reflect integration of communication and sensing channels.
\end{itemize}

\begin{itemize}
\item[2)] 
In many existing works of ISAC, communication and sensing channels are generated independently. In \cite{Ref21}, modulation of ISAC system are optimized based on information transmission rates. However, it does not distinguish between communication channels and sensing channels, both of which are considered to be frequency selective channels. In \cite{Ref22}, a vehicle ISAC system based on IEEE 802.11ad is presented, where communication channel is considered as a Rice channel and sensing channel is considered as a path-based model. Similarly, communication channel and sensing channel are considered as a Rayleigh model and a path-based model respectively in \cite{Ref23} to design a dual function radar communication system. In this case, communication and sensing performance of ISAC system can only be evaluated independently. Some researches use the same model for communication and sensing channels. For instance, both communication and sensing channels are considered as path-based or cluster-based channel models in \cite{Ref24} and \cite{Ref25}, respectively. However, for these models, correlation between communication and sensing channels is not well reflected. 
\end{itemize}

\begin{itemize}
\item[3)]
In a few existing researches, correlatively modeling between communication and sensing channels has attracted some attention. In \cite{Ref26}, a millimeter-wave ISAC system is designed, in which communication clusters are assumed as a part of sensing channel. In \cite{Ref27}, a multi-link channel architecture in ISAC scenarios is proposed. However, these ISAC channel models lack actual measurement validation. Ref.\cite{Ref28} analyzes spatial consistency characteristics between communication and sensing channels based on measurements with a vehicle-mounted radar, which only focuses on power spectrum analysis. In \cite{Ref29}, communication and sensing indoor channel measurement is conducted at 28 GHz, capturing shared features between communication and sensing channels. However, these ISAC channel models are difficult to integrate with the existing standard channel models. 
Moreover, deterministic channel models such as Ray-tracing (RT) have ability to provide the correlation modeling of ISAC channels. For instance, Ref.\cite{Ref30} and \cite{Ref31} analyze ISAC channels in indoor static scenarios and outdoor vehicle-to-vehicle scenarios respectively based on RT models. In addition, a deterministic ISAC channel model is proposed based on indoor millimeter-wave geometric structures in \cite{Ref32}. In \cite{Ref33}, ISAC channel corresponding to target scatterer is proposed based on points models, and sensing characteristics of each point in scatterer are analyzed. However, these deterministic models have high computational cost and complex implementation. To sum up, highlighting correlation between communication and sensing channels to evaluate ISAC systems remains a crucial research focus.
\end{itemize}

\subsection{Major Contributions}

As discussed before, based on the current standard channel models such as 3GPP TR 38.901\cite{Ref34}, deriving the corresponding sensing channels under the same procedure to integrate ISAC channels is a reasonable and compatible modeling approach. In order to fill the aforementioned gaps, this paper proposes a cluster-based ISAC statistical channel model. Note that the ISAC channel measurements are limited to the specific environment. The work in this paper shows an innovative correlatively sensing and communication channel modeling approach, in which model parameters are extracted from actual measurements and a generalized statistical model is established. The work can provide guidance for channel measurement and modeling in both the current and other scenarios, and the main contributions are summarized as follows:

\begin{itemize}
\item[1)]
A new framework of ISAC channel model based on 3GPP standard is proposed. Clusters are divided into two categories, communication clusters and sensing clusters, which are used to describe multi-path components (MPCs) statistical distribution of power, delay, angle, etc.
\end{itemize}

\begin{itemize}
\item[2)]
A wideband MIMO channel measurement system is used to conduct directional ISAC channel measurements at 28 GHz. Both communication channels and sensing channels are measured on the consistent outdoor environment. 
\end{itemize}

\begin{itemize}
\item[3)]
We extract distribution of clusters in ISAC channel. Some novel cluster types, including shared sensing cluster and newborn sensing cluster, are introduced to specialize sensing channels and model correlation between communication and sensing channels.
\end{itemize}

\begin{itemize}
\item[4)]
The proposed channel model is validated by comparing with measurement data, which guarantees accuracy of the proposed model and the related analysis in the paper.
\end{itemize}

The remainder of this paper is structured as follows. In Section II, a cluster-based statistical channel model is proposed for ISAC scenarios. Section III introduces ISAC channel measurement system and measurement campaign. In Section IV, measurement data processing and modeling for sensing clusters are presented. Section V presents ISAC channel model implementation and validation. Finally, conclusion is given in Section VI.

\section{ISAC Channel model}
It is widely known that MPCs are generally distributed in groups (clusters)\cite{Ref35,Ref36} and a cluster-based model has less complex when keeping the accuracy\cite{Ref37}. In this section, a cluster-based ISAC statistical channel model is proposed to capture correlation between communication and sensing channels. In addition, a framework of ISAC channel modeling is proposed based on the structure of 3GPP channel model. We will highlight the required extension of ISAC channel model based on the procedure in 3GPP TR 38.901. \cite{Ref34}.

\subsection{Cluster-Based ISAC Channel}

\begin{figure}[t]
    \centering
        \includegraphics[width=1\linewidth]{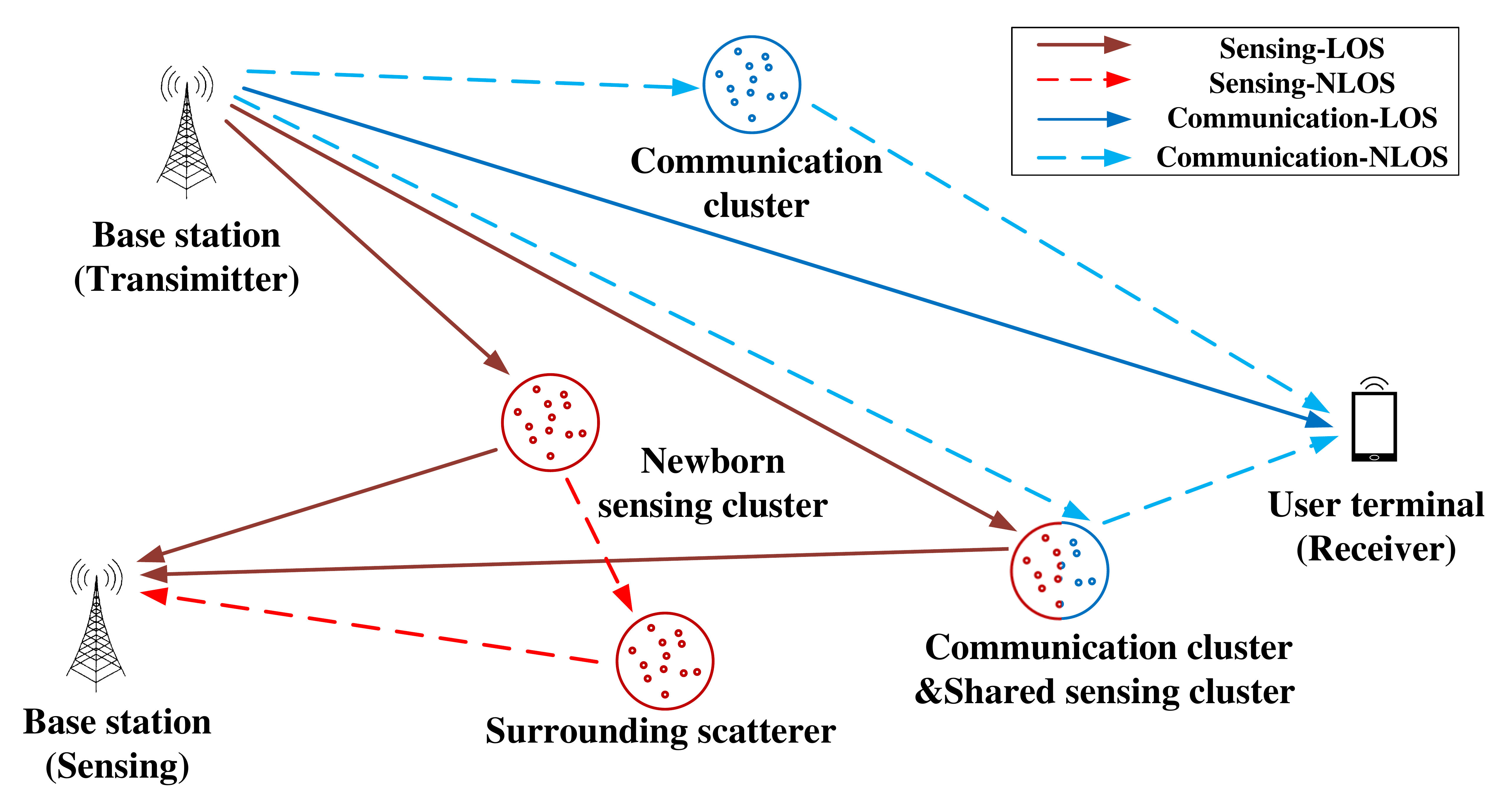}
    \caption{Illustration of cluster-based ISAC channels.}
    \label{fig1}
\end{figure}

This paper considers a 3D cluster-based ISAC channel as shown in Fig. 1. Scatterers are randomly distributed in environment, following spatial characteristics between communication and sensing channels. The clusters contribute to communication channels or sensing channels are considered as communication clusters or sensing clusters, respectively. As shown in Fig. 1, communication and sensing channels are illustrated by blue and red lines, respectively. More detailed introductions about Fig. 1 are as follows:

$\bullet$ Base station (BS) and user terminal (UT) are also called as transmitting terminal (TX), sensing terminal (SX) and receiving terminal (RX) in this paper. When SX is located on TX, it is considered as a “monostatic” ISAC channel. Otherwise, it is a “bistatic” ISAC channel, which is shown in the figure.

$\bullet$ If sensing echoes only have one interaction with scatterers (e.g., perception target), propagation condition is considered as Line-of-Sight (LOS) sensing propagation, otherwise it is considered as Non-Line-of-Sight (NLOS) sensing propagation, marked by solid and dashed lines in the figure respectively. For LOS sensing propagation, amplitude, phase, and other characteristics of echoes are determined by target scatterers, resulting in effective sensing signals. For NLOS sensing propagation, amplitude, phase, and other characteristics of echoes are jointly impacted by multiple scatterers, resulting in multiple-bounce and invalid sensing signals.

$\bullet$ Clusters in LOS sensing propagation mainly include two types: shared sensing clusters and newborn sensing clusters. Shared sensing clusters are defined that clusters in communication channels still exist in the corresponding sensing channels and contribute to communication and sensing channels simultaneously; while newborn sensing clusters only contribute to sensing channels.

In the proposed cluster-based ISAC channel, communication and sensing clusters from the same physical scatterer share inter-cluster parameters, such as arrival angles, departure angles, spatial position, etc., while communication and sensing clusters from different physical scatterers are independent to each other. Besides, in order to better represent physical characteristics of scatterers, sensing clusters are expected to have more intra-cluster parameters than communication clusters, such as radar cross section (RCS). In addition, when multiple BS and UT are deployed with same environment in simulation, sensing clusters in different BS-UT links interact with each other, making it possible for one BS to perceive scatterers in other BS-UT links.

\subsection{Generation Procedure}
The aim of proposed modeling approach is to derive sensing channels from 3GPP standard communication channel model, which is suitable to target detection, localization, tracking, sensing assisted communication, communication assisted sensing, etc. 
\begin{figure*}[t]
    \centering
        \includegraphics[width=1\linewidth]{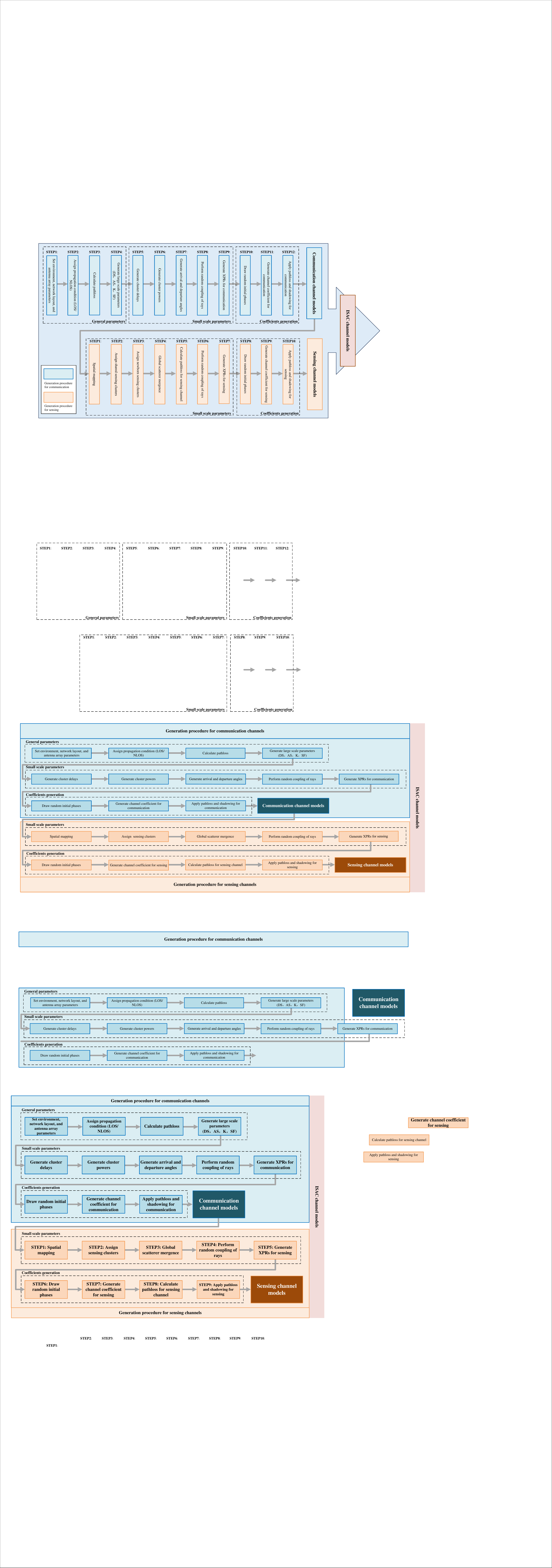}
    \caption{Framework for ISAC channel modeling.}
    \label{fig1}
\end{figure*}
Based on 3GPP TR 38.901, generation procedure for ISAC channel modeling is shown in Fig. 2. The blue highlighted fields represent the same steps with 3GPP standard channels, which is used to develop communication channel model. While the yellow highlighted fields represent extension parts for sensing channel. With the proposed framework, communication channel and sensing channel are generated using the same UE parameters, which can be divided into three parts.

\subsubsection{General Parameters}

including UE parameters and large-scale parameter (LSP) generation. At first, UE parameters are fed into procedure, such as types of scenario (urban micro, urban macro, rural macro, etc.), network layout (numbers and 3D locations of BS and UT, speeds and directions of UTs, frequency and bandwidth) and antenna array (antenna field patterns and array geometries of BSs and UTs). Based on UE parameters, propagation condition (LOS/NLOS) can be assigned for different BS-UT links. Afterwards, pathloss of communication channel and LSPs are generated.

\subsubsection{Small Scale Parameters}

including cluster-based propagation parameters. For communication channels, small scale parameters (SSPs) of each individual cluster and rays within cluster are generated based on the specific LSPs in general parameters and the predefined statistical models in 3GPP. Then perform random coupling of rays and apply cross polarization power ratios (XPR) to clusters. For sensing channels, SSPs are obtained based on the generated communication channels, in which location coordinates of clusters can be mapped according to delays, powers and angles. Then assign sensing clusters, perform random coupling of rays and apply cross XPRs to clusters.

\subsubsection{Coefficients Generation}

including channel impulse response. Based on the generated SSPs, channel coefficient in time domain can be generated based on field patterns of receiver antenna and transmitter antenna respectively. Then coefficients of rays are added together to generate channel coefficients, and pathloss as well as shadowing are applied on them. Note that pathloss for sensing channel is calculated separately for each clusters due to different propagation length.

\subsection{Extension for Sensing Channels}

In this section, we focus on extension procedure for sensing channels, especially on spatial mapping module (STEP1), assign sensing clusters module (STEP2), global scatterer mergence module (STEP3), channel coefficient generation module (STEP7) and pathloss calculation module (STEP8). 

\subsubsection{Spatial Mapping}
Based on generated communication cluster delays $\tau$, powers $P$, zenith angle of arrival $\theta_{ZOA}$, azimuth angle of arrival $\phi_{AOA}$, zenith angle of departure $\theta_{ZOD}$ and azimuth angle of departure $\phi_{AOD}$, location of cluster can be calculated\cite{Ref38}. Fig. 3 illustrates spatial mapping of clusters based on 3GPP standard channel. 

\begin{figure}[t]
    \centering
        \includegraphics[width=0.9\linewidth]{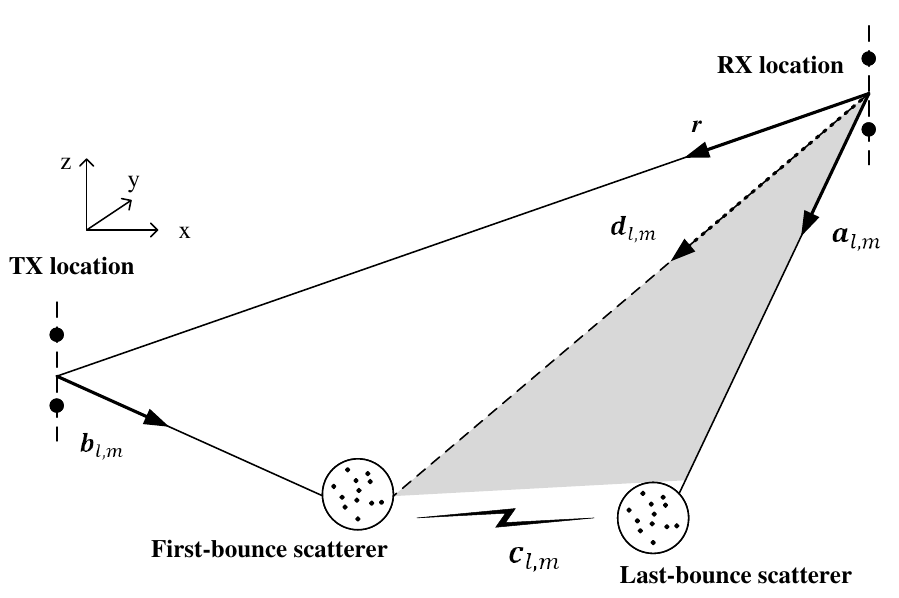}
    \caption{Spatial mapping illustration of clusters.}
    \label{fig1}
\end{figure}

First-bounce scatterer (FBS) and last-bounce scatterer (LBS) correspond to the first and last scatterers in radio propagation respectively. For subpath $l$ in communication cluster $m$, $\boldsymbol{b}_{l, m}$ represents vector pointing from TX to FBS, $\left|\boldsymbol{c}_{l, m}\right|$ represents length of path from FBS to LBS, $\boldsymbol{a}_{l, m}$ represents vector from RX to LBS. Assuming delay of LOS path is zero, thus total propagation length of subpath $l$ is follows:
\begin{equation}
d_l=\left|\boldsymbol{b}_{l, m}\right|+\left|\boldsymbol{c}_{l, m}\right|+\left|\boldsymbol{a}_{l, m}\right|=\tau_l \cdot c+|\boldsymbol{r}|
\end{equation}
where $|\boldsymbol{r}|$ is distance between TX and RX. Direction vector of path is calculated as
\begin{equation}
\widehat{\boldsymbol{a}}_{l, m}=\left(\begin{array}{c}
\cos \phi_{l, m, A O A} \cdot \sin \theta_{l, m, Z O A} \\
\sin \phi_{l, m, A O A} \cdot \sin \theta_{l, m, Z O A} \\
\cos \theta_{l, m, Z O A}
\end{array}\right)=\frac{\boldsymbol{a}_{l, m}}{\left|\boldsymbol{a}_{l, m}\right|}
\end{equation}
\begin{equation}
\widehat{\boldsymbol{b}}_{l, m}=\left(\begin{array}{c}
\cos \phi_{l, m, A O D} \cdot \sin \theta_{l, m, Z O D} \\
\sin \phi_{l, m, A O D} \cdot \sin \theta_{l, m, Z O D} \\
\cos \theta_{l, m, Z O D}
\end{array}\right)=\frac{\boldsymbol{b}_{l, m}}{\left|\boldsymbol{b}_{l, m}\right|}
\end{equation}
In order to ensure that spatial mapping is reasonable, $d_ {min}$ is introduced as distance between TX or RX and the nearest scatterer, and making $\left|\boldsymbol{b}_{l, m}\right| \sim U\left(d_{\min }, d_l / 2\right)$. For triangular structure composed of RX, FBS, and LBS (shaded in Fig. 3), distance between RX and LBS $\left|\boldsymbol{a}_{l, m}\right|$ can be calculated according to cosine theorem, as follows: 
\begin{equation}
\left|\boldsymbol{a}_{l, m}\right|=\frac{\left(d_l^{\prime}\right)^2-\left|\boldsymbol{d}_{l, m}\right|^2}{2 \cdot\left(d_l^{\prime}-\boldsymbol{d}_{l, m}{ }^T \widehat{\boldsymbol{a}}_{l, m}\right)}
\end{equation}
where $\boldsymbol{d}_{l, m}$ and $d_l^{\prime}$ represent vector and propagation length from RX to FBS respectively, expressed as follows:
\begin{equation}
d_l^{\prime}=d_l-\left|\boldsymbol{b}_{l, m}\right|
\end{equation}
\begin{equation}
\boldsymbol{d}_{l, m}=\boldsymbol{r}+\widehat{\boldsymbol{b}}_{l, m} \cdot\left|\boldsymbol{b}_{l, m}\right|
\end{equation}
 In fact, $\left|\boldsymbol{a}_{l, m}\right|$ might be smaller than $d_ {min}$. This case is treated by setting $\left|\boldsymbol{c}_{l, m}\right|=0$ and calculating new location based on delay and departure angle. It often occurs when propagation delay is short, that is, path length is only slightly longer than LOS path. Assuming TX is located at the origin of global coordinate system, then location coordinates $\left[\boldsymbol{X}, \boldsymbol{Y}, \boldsymbol{Z}\right]$ of FBS and LBS are mapping as follows:

\begin{equation}
\left[\boldsymbol{X}_{FBS}, \boldsymbol{Y}_{FBS}, \boldsymbol{Z}_{FBS}\right]=\widehat{\boldsymbol{b}}_{n, m} \cdot\left|\boldsymbol{b}_{n, m}\right|
\end{equation}

\begin{equation}
\left[\boldsymbol{X}_{LBS}, \boldsymbol{Y}_{LBS}, \boldsymbol{Z}_{LBS}\right]=-\boldsymbol{r}+\widehat{\boldsymbol{a}}_{n, m} \cdot\left|\boldsymbol{a}_{n, m}\right|
\end{equation}

\subsubsection{Assign Sensing Clusters}

Shared sensing clusters are generated based on spatial mapping of communication clusters. In the proposed ISAC channels, there exists communication clusters with their own spatial mapping. Some of them only contribute to communication channels while some of them are shared with sensing channels. Whether communication clusters are considered as shared sensing clusters can be evaluated in term of an evolution probability\cite{Ref39}. A higher probability indicates that more clusters in communication channels are shared with sensing channels. Note that if distance between SX and target is small, physical scatterers are
more easily perceived by SX, and evolution probability is higher.

Newborn sensing clusters are generated independently. According to measured statistical distribution, procedure of communication clusters is employed to generate newborn sensing clusters with their own spatial mapping.  Note that UT is also modeled as a newborn sensing cluster if communication channel is LOS propagation.

\subsubsection{Global Scatterer Mergence}
It is obvious that number of sensing clusters in ISAC channel is related to distribution of scatterers in surrounding environment. Therefore, it is essential to limit global number of sensing clusters to avoid explosive sensing clusters in process of generation. If redundant sensing clusters in simulation space are generated, clusters that are closer to each other and merged into one scatterer by agglomerative hierarchical algorithm \cite{Ref40}, meaning that they correspond to the same physical target, so as to ensure that number of global sensing clusters in ISAC channel is finite and stationary. The inter-cluster similarity $L(r,s)$ are evaluated in term of spatial position, which is defined as the average distance in coordinates of all sample pairs within two clusters $r$ and $s$, calculated as follows:
\begin{equation}
L(r, s)=\frac{1}{n_r n_s} \sum_{i=1}^{n_r} \sum_{j=1}^{n_s} \left(x_{r,i}- x_{s,j}\right)^2
\end{equation}
where $n_r$ and $n_s$ represent number of samples within clusters, and $x_{r,i}$ and $x_{s,j}$ represent positional coordinates of each sample respectively.

\subsubsection{Generate Channel Coefficient for Sensing}
 
Considering all generated sensing clusters as perception targets in environment, 3GPP standard procedure is employed for each perception target to generate sensing channel coefficients. According to distance from TX to target, and from target to SX, sensing propagation condition can be assigned based on LOS probability of 3GPP TR38.901. Patterns of sensing propagation are shown in Table I. Only in the case that TX to target and target to SX are both LOS conditions, the propagation condition is LOS sensing. 

\begin{table}[t]
\centering
\caption{Patterns of sensing propagation.} \label{Table1Label}
\scalebox{1.1}{
\begin{tabular}{ccc}
\hline
Propagation conditions & TX to Target & Target to SX \\ \hline
LOS sensing            & LOS           & LOS           \\
NLOS sensing           & LOS           & NLOS          \\
NLOS sensing           & NLOS          & LOS           \\
NLOS sensing           & NLOS          & NLOS          \\ \hline
\end{tabular}}
\end{table}

Therefore, channel coefficient of LOS sensing path and NLOS sensing path for each sensing and transmitter element pair $u$,$s$ are given respectively as in (10) and (11),
\begin{figure*}[!t]

\begin{equation}
\begin{gathered}
h_{u, s}^{L O S}(t)=\left[\begin{array}{l}
F_{s x, u, \theta}\left(\theta_{L O S, Z O A}, \phi_{L O S, A O A}\right) \\
F_{s x, u, \phi}\left(\theta_{L O S, Z O A}, \phi_{L O S, A O A}\right)
\end{array}\right]^T\left[\begin{array}{cc}
1 & 0 \\
0 & -1
\end{array}\right]\left[\begin{array}{l}
F_{t x, s, \theta}\left(\theta_{L O S, Z O D}, \phi_{L O S, A O D}\right) \\
F_{t x, s, \phi}\left(\theta_{L O S, Z O D}, \phi_{L O S, A O D}\right)
\end{array}\right] \\
\exp \left(\frac{j 2 \pi\left(\hat{r}_{t x, l, m}^T \cdot \bar{d}_{t x, s}\right)}{\lambda_0}\right) \exp \left(\frac{j 2 \pi\left(\hat{r}_{s x, l, m}^T \cdot \bar{d}_{s x, u}\right)}{\lambda_0}\right) \exp \left(\frac{j 2 \pi\left(\left(\hat{r}_{t x, l, m}^T+\hat{r}_{r x, l, m}^T\right) \cdot \bar{v}\right) t}{\lambda_0}\right)
\end{gathered}
\end{equation}

\begin{equation}
\begin{aligned}
& h_{u, s, l, m}^{N L O S}(t) \\
& =\left[\begin{array}{l}
F_{s x, u, \theta}\left(\theta_{l, m, Z O A}, \phi_{l, m, A O A}\right) \\
F_{s x, u, \phi}\left(\theta_{l, m, Z O A}, \phi_{l, m, A O A}\right)
\end{array}\right]^T\left[\begin{array}{cc}
\exp \left(j \Phi_{l, m}^{\theta \theta}\right) & \sqrt{\kappa_{l, m}^{-1}} \exp \left(j \Phi_{l, m}^{\theta \phi}\right) \\
\sqrt{\kappa_{l, m}^{-1}} \exp \left(j \Phi_{l, m}^{\phi \theta}\right) & \exp \left(j \Phi_{l, m}^{\phi \phi}\right)
\end{array}\right]\left[\begin{array}{l}
F_{t x, s, \theta}\left(\theta_{l, m, Z O D}, \phi_{l, m, A O D}\right) \\
F_{t x, s, \phi}\left(\theta_{l, m, Z O D}, \phi_{l, m, A O D}\right)
\end{array}\right] \\
& \quad \exp \left(\frac{j 2 \pi\left(\hat{r}_{t x, l, m}^T \cdot \bar{d}_{t x, s}\right)}{\lambda_0}\right) \exp \left(\frac{j 2 \pi\left(\hat{r}_{s x, l, m}^T \cdot \bar{d}_{s x, u}\right)}{\lambda_0}\right) \exp \left(\frac{j 2 \pi\left(\left(\hat{r}_{t x, l, m}^T+\hat{r}_{s x, l, m}^T\right) \cdot \bar{v}\right) t}{\lambda_0}\right)
\end{aligned}
\end{equation}
\end{figure*}
where $F_{sx,u,\theta}$ and $F_{sx,u,\phi}$ are field patterns of sensing antenna element $u$ in direction of spherical basis vectors, $\theta$ and $\phi$ respectively. $F_{tx,s,\theta}$ and $F_{tx,s,\phi}$ are field patterns of sensing antenna element $s$ in direction of spherical basis vectors, $\theta$ and $\phi$ respectively. $\hat{r}_{sx, l, m}$ and $\hat{r}_{tx, l, m}$ are spherical unit vector with arrival angle and departure angle respectively. $\bar{d}_{sx, u}$ and $\bar{d}_{t x, s}$ are location vectors of sensing antenna element $u$ and transmit antenna element $s$ respectively. $\kappa_{l, m}$ is cross polarisation power ratio in linear scale and $\lambda_0$ is wavelength. ${\Phi^{\theta \theta}, \Phi^{\theta \phi}, \Phi^{\phi \theta}, \Phi^{\theta \theta}}$ are random initial phases. $\bar{v}$ is target velocity vector.

\subsubsection{Calculate Pathloss for Sensing}
Unlike pathloss calculation in communication channels, pathloss of sensing channels is calculated separately for each clusters due to different propagation length. In free space propagation, pathloss of communication channel with omnidirectional isotropic antennas is given as follows:
\begin{equation}
P L_{\text {com }}(d)=\frac{(4 \pi)^2 d^2}{\lambda^2}
\end{equation}
where $d$ is distance between TX and RX. Correspondingly, pathloss of sensing channel is given as follows:
\begin{equation}
P L_{\text {sen}}\left(d_1, d_2, \sigma_{R C S}\right)=\frac{64 \pi^3 d_1^2 d_2^2}{\lambda^2 \sigma_{R C S}}
\end{equation}
where $d_1$, $d_2$ and $\sigma_{R C S}$ are distances between TX and target, target and SX, RCS of target, respectively. Based on (12) and (13), pathloss of sensing channels is given as follows:
\begin{equation}
\begin{gathered}
P L_{\text {sen }}\left(d_1, d_2, \sigma_{R C S}\right) [dB] =\\P L\left(d_1\right)+P L\left(d_2\right)-10 \log \sigma_{R C S}+10 \log \frac{\lambda^2}{4 \pi} 
\end{gathered}
\end{equation}
where pathloss of sensing channels can be calibrated based on the widely used communication pathloss model from 3GPP TR38.901, expressed as $P L(d)$ in (14). 
At last, apply pathloss and shadowing for communication channels and sensing channels and integrate them as ISAC channels.

\section{Measurement campaign}

\subsection{Measurement System}

A wideband MIMO measurement system is used for ISAC channel measurements, including signal generator and signal analyzer, up-conversion and down-conversion modules, reference clock modules, horn and array antennas, and electronic switch (integrated within down-conversion module). The National Instruments (NI) PXIe-5745, also known as PXI FlexRIO Signal Generator, is used as signal generator. The NI PXIe-5775, also known as  PXI FlexRIO Digitizer, is used as signal analyzer \cite{Ref41}. Sounding signal employed in measurements is a multi-carrier signal with 1 GHz bandwidth and frequency of 28 GHz. A power amplifier with a gain of 28 dB is integrated within up-conversion module. Both TX and RX employ a global positioning system (GPS)-tamed Rubidium clock to achieve synchronization, which can provide a 10 MHz reference clock impulse. The detailed configurations of measurement system are provided in Table II.
\begin{table}[]
\centering
\caption{Configurations of ISAC measurement system.} \label{Table1Label}
\scalebox{1.2}{
\begin{tabular}{cc}
\hline
Parameters                 & Value                \\ \hline
Center frequency           & 28 GHz               \\
Bandwidth                  & 1 GHz                \\
Transmit power             & 28 dBm               \\
Range resolution           & 0.3 m                 \\
Sounding signal            & Multi-carrier signal \\
Number of frequency points & 1024                 \\
Sample rate                & 6.4 GHz              \\
Transmitter antenna        & Horn antenna         \\
Sensing antenna            & 4$\times$8 array antenna    \\
Receiver antenna           & 4$\times$8 array antenna    \\ \hline
\end{tabular}}
\end{table}

\begin{figure}[t]
\centering
\subfigure[]{\includegraphics[width=1.7in]{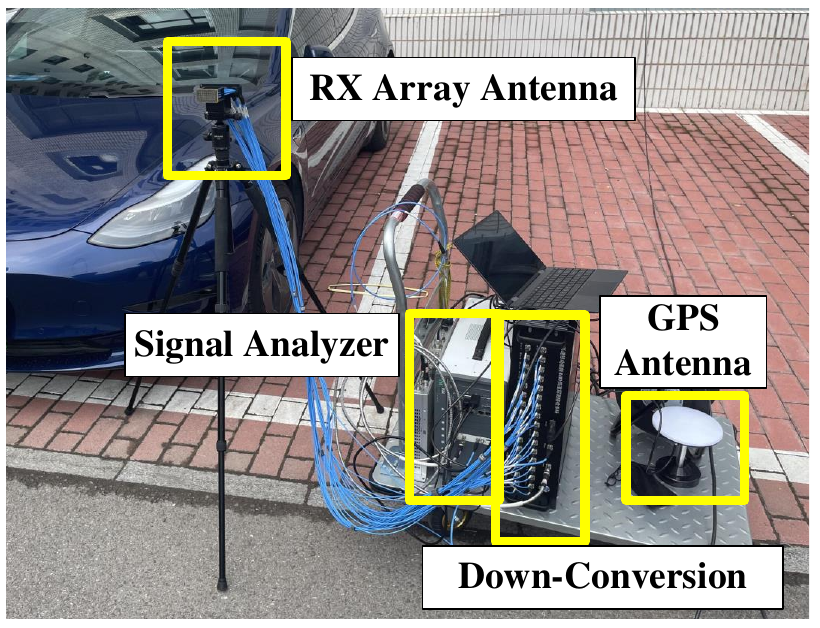}}
\subfigure[]{\includegraphics[width=1.7in]{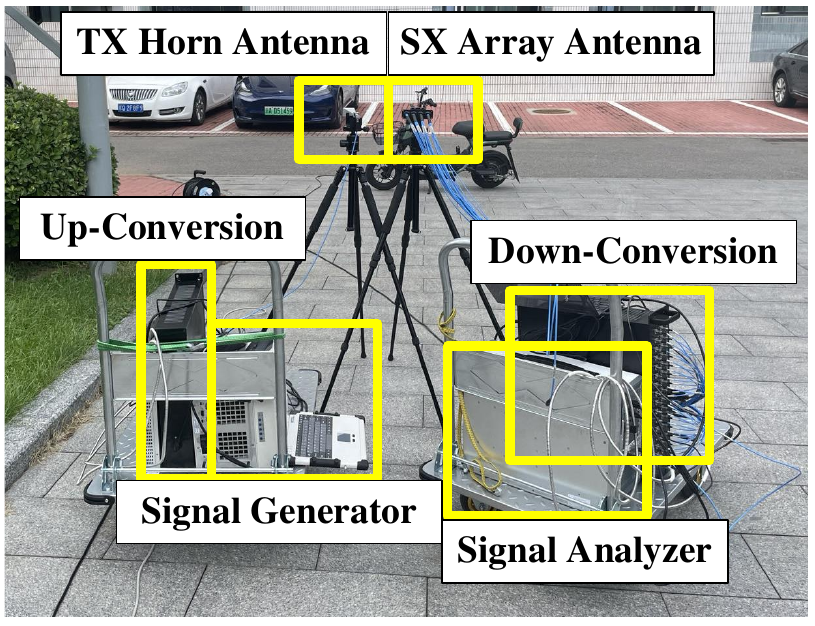}}
\caption{ISAC channel measurement system. (a) RX. (b) TX and SX.}
\end{figure}

As shown in Fig. 4, in TX, baseband signal is generated by signal generator and then is up-converted to millimeter wave through up-conversion module. After amplification by power amplifier, sounding signal is transmitted through a horn antenna with a gain of 20 dB. The whole process is synchronized by a rubidium atomic reference clock.
In RX, a 4×8 antenna array is employed to receive sounding signal and send it to down-conversion module through a high-speed electronic switch. The signal is then down-converted to baseband and stored in signal analyzer. The whole process is synchronized by a rubidium atomic reference clock, which is calibrated with that of TX through GPS.
In SX, it is located close to TX, enabling it to share rubidium atomic clock for synchronization. Signal receiving process in SX is the same as RX, thus no further explanation is presented due to space limitation.

\subsection{Calibration}

\begin{figure}[t]
\centering
\subfigure[]{\includegraphics[width=1.74in]{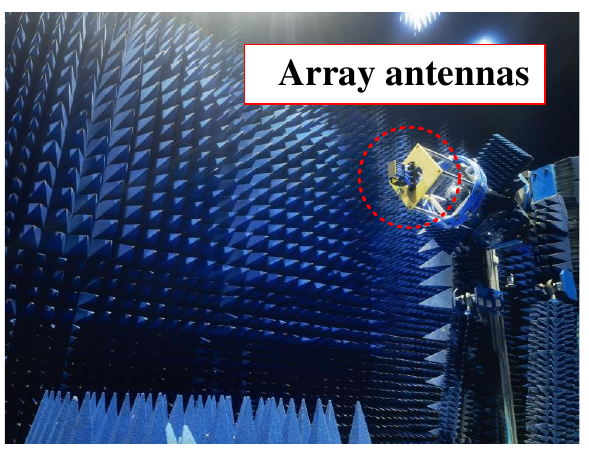}}
\subfigure[]{\includegraphics[width=1.7in]{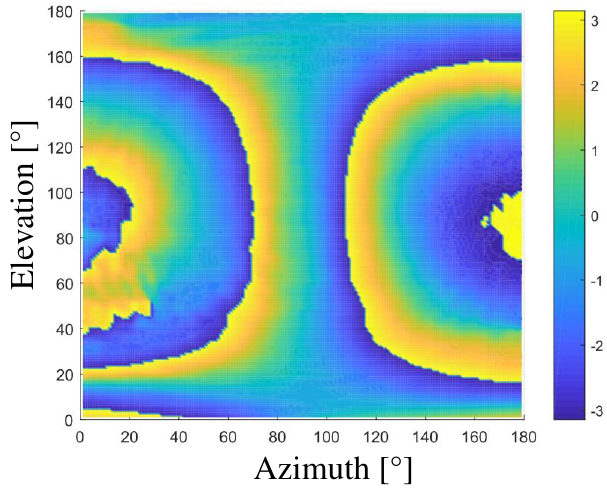}}

\caption{Illustration of antenna calibration. (a) Array calibration in anechoic chamber. (b) Measured phase radiation pattern of element 16 at 28 GHz as an example.}
\end{figure}

Measurement system calibration is a crucial step in ensuring accurate results. It comprises of two fundamental parts: back-to-back measurement and antenna calibration. Back-to-back calibration aims to eliminate impact of measurement systems, including cables, switches, transceivers and etc. Besides, in order to obtain accurate measurement-based MPCs estimation, antenna calibration is also essential \cite{Ref42,Ref43}. In this paper, measurements of radiation patterns of each array antenna element are conducted in an anechoic chamber, as shown in Fig. 5(a). For the array, we measured three-dimensional gain and phase radiation pattern in every spatial direction of front, ranging from azimuth 1-180 degrees and elevation angle 1-180 degrees. Fig. 5(b) displays a three-dimensional phase radiation pattern of element 16 of the array at 28 GHz. The obtained calibration data will be used as steering vectors for MPC estimation. 

\subsection{ISAC Channel Measurements}

\subsubsection{Dynamic Measurement}
\begin{figure}[t]
\centering
\subfigure[]{\includegraphics[width=3.5in]{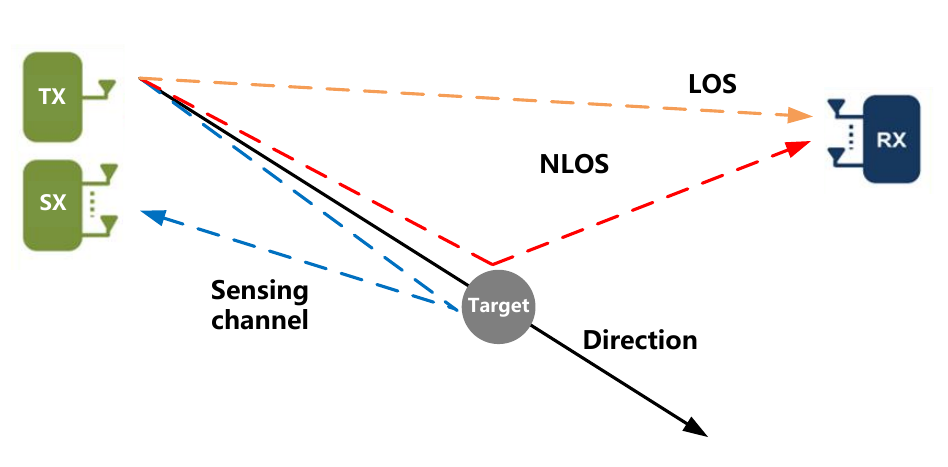}}
\subfigure[]{\includegraphics[width=3.5in]{Fig6b.pdf}}
\caption{Dynamic measurement scenario for ISAC channel. (a) Measurement campaign. (b) Scenario photo.}
\end{figure}
Dynamic measurement scenario is illustrated in Fig. 6(a), which aims to obtain dynamic variation from the same target cluster in communication and sensing channels. The measured data can be further used to obtain statistical models of evolution probability to determine shared sensing clusters. A moving scatterer is manually designed to move along the pre-set route with stable velocity, and dynamic variation of clusters are simultaneously measured in both communication and sensing channels. 
Measurements are carried out in the playground of Beijing Jiaotong University, as shown in Fig. 6(b). TX and SX, which transmit and receive signals along the direction of target movement, are about 5 meters high, and marked as starting point of movement route. RX is 1.5 meters high and has a LOS distance of 22 meters from TX. During measurements, a pedestrian is used as the moving target scatterer, who moves at a constant speed along a fixed route with a limited length of approximately 50 meters.

\subsubsection{Static Measurement}

As shown in Fig. 7(a), static measurement is conducted to obtain clusters with random scatterer distribution in communication and sensing channels, which can be further used to obtain relationship between newborn sensing clusters and communication clusters. Communication channel is measured at different positions, and the corresponding sensing channels under same environment is measured simultaneously. In Fig. 7(a), the red objects, which are marked as new scatterers, represent scatterers that only contribute to sensing channels and result in newborn sensing clusters.
Measurements are conducted in a park of Beijing Jiaotong University, as shown in Fig 7(b). Some common scatters are randomly distributed in spatial environment, including trees, lawns, street lights, cars and buildings. TX and SX are mounted at a height of approximately 5 meters. RX is mounted with about 1.5 meters high. The whole measurement includes 14 groups of communication channels (having LOS distances of 11.3, 15.7, 18.1, 17.1, 23.7, 27.2, 14.6, 14.1, 15.3, 16.2, 10.8, 9.2, 23.7, 18.2 meters respectively) and two groups of sensing channels (having two different directions of TX antenna respectively).

\section{data processing and modeling}

\subsection{Shared Sensing Clusters Evolution}

In the proposed models, in order to model correlation between communication and sensing channels, shared sensing clusters are defined that a cluster in communication channel still contributes to the corresponding sensing channel. Each cluster in communication channel has potential to be considered as shared sensing cluster. The evaluation of whether clusters in communication channels are shared with sensing channels can be determined through an evolution probability, i.e.,$P_{\text{evol}}$. 

A higher $P_{\text{evol}}$ means that a cluster in communication channel is more easily evolved to a shared sensing cluster. When applying $P_{\text{evol}}$ to each cluster, a higher $P_{\text{evol}}$ results in more clusters in the communication channel being shared with the sensing channel. In this paper, $P_{\text{evol}}$ is introduced as the ratio of number of multipaths corresponding to one target scatterer in communication and sensing channels, which is expressed as follows:
\begin{equation}
P_{\text{evol}}=\frac{N_{\text{sen}}}{N_{\text{com}}}
\end{equation}
where $N_{\text{sen}}$ and ${N_{\text{com}}}$ denote numbers of multipaths from the target scatterer in sensing and communication channels respectively. When $N_{\text{sen}}<{N_{\text{com}}}$, $P_{\text{evol}}$ is less than 1 and it means that part of clusters in communication channel are shared with sensing channel. When $N_{\text{sen}} \ge {N_{\text{com}}}$, $P_{\text{evol}}$ is greater than 1. In this case, all clusters in communication channel are shared with sensing channel, and $P_{\text{evol}}$ is adjusted to 1.

\begin{figure}[t]
\centering
\subfigure[]{\includegraphics[width=3.5in]{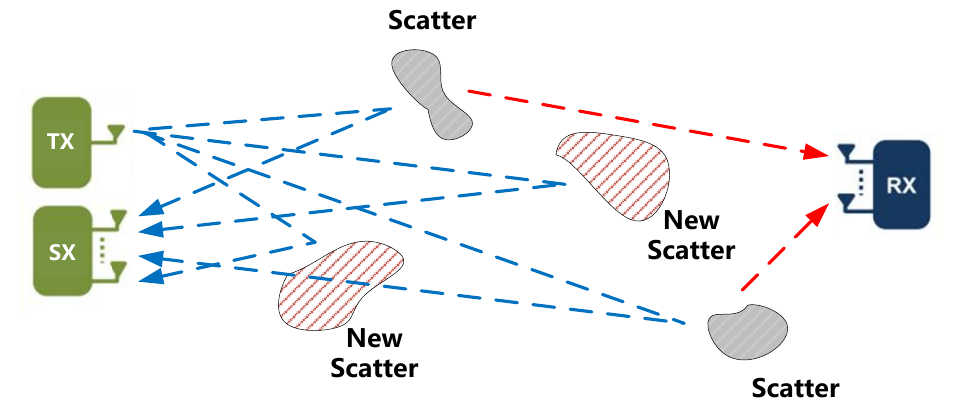}}
\subfigure[]{\includegraphics[width=3.5in]{Fig7b.pdf}}

\caption{Static measurement scenario for ISAC channel. (a) Measurement campaign. (b) Scenario photo.}
\end{figure}

\begin{figure}[t]
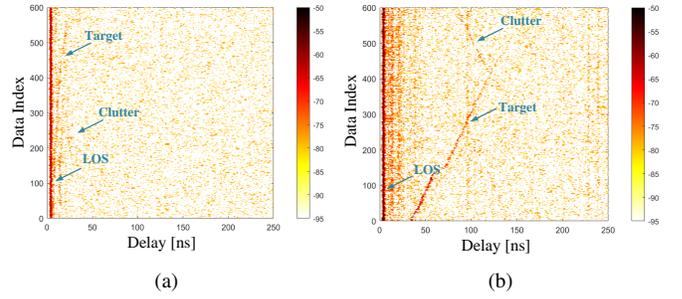

\centering
\subfigure[]{\includegraphics[width=1.7in]{Fig8a.pdf}}
\subfigure[]{\includegraphics[width=1.7in]{Fig8b.pdf}}
\caption{Measured PDPs in dynamic measurement scenario. (a) Communication channel. (b) Sensing channel.}
\end{figure}

Shared sensing cluster is highly related to distance between SX and scatterers. The distance is smaller, scatterers are more easily perceived by SX, while the distance is large, scatterers are more difficult to be perceived by SX. Although there also exist newborn sensing clusters in ISAC channels, which contribute only to sensing channels, they are not considered in this section. Therefore, $P_{\text{evol}}$ is negatively correlated with perception distance (between scatter and SX, marked as $r$) and negatively correlated with propagation length (from TX to RX, marked as $L$). A heuristic statistical model is proposed for $P_{\text{evol}}$ as follows:
\begin{equation}
P_{\text{evol}}(\bar{r})=a \cdot e^{b \cdot \bar{r}}
\end{equation}
where $a$ and $b$ are fitted parameters according to measurements. Noted that input of model is normalized by distance $d$ between TX and RX in communication channel, which is $\bar{r}=\frac{r}{d} \cdot \frac{L}{d}$. As the normalized distance $\bar{r}$ becomes larger, $P_{\text{evol}}$ is infinitely close to zero, which means the clusters with this far distance is impossible to be perceived and share with sensing channel. If the normalized distance $\bar{r}$ is very small, resulting a probability $P_{\text{evol}}$ of 1, then the clusters in this distance can be perceived totally.

Power-delay profiles (PDPs) of communication and sensing channels in dynamic measurement scenario are presented in Fig. 8, and it can be observed that movement trajectory of target scatterer is obvious in PDPs, marked as ‘Target’. Besides, during the measurements, multipath from non-target objects are unavoidable to be collected simultaneously, which are marked as ‘Clutter’. Noted that only the multipaths corresponding to target are used for modeling and analysis, and clutter component is eliminated. Due to the continuous changes in position of moving target, its corresponding multipath is also continuously changing. To accurately extract the multipaths associated with moving target, we employ a dynamic multipath tracking algorithm to separate and extract multipath trajectory in PDP, thereby identifying the multipaths corresponding to moving target, as illustrated in Fig. 9. Subsequently, multipath parameters are jointly estimated using the SAGE algorithm, which iteratively computes the maximum likelihood estimation of the MPCs parameters such as powers, delays, and angles\cite{Ref44}. Based on SAGE estimation results, including estimated power, delay and angle information, number of multipaths corresponding to each delay bin is counted and the multipaths that match with tracking trajectory in PDP are separated and identified.

\begin{figure}[t]
    \centering
        \includegraphics[width=1\linewidth]{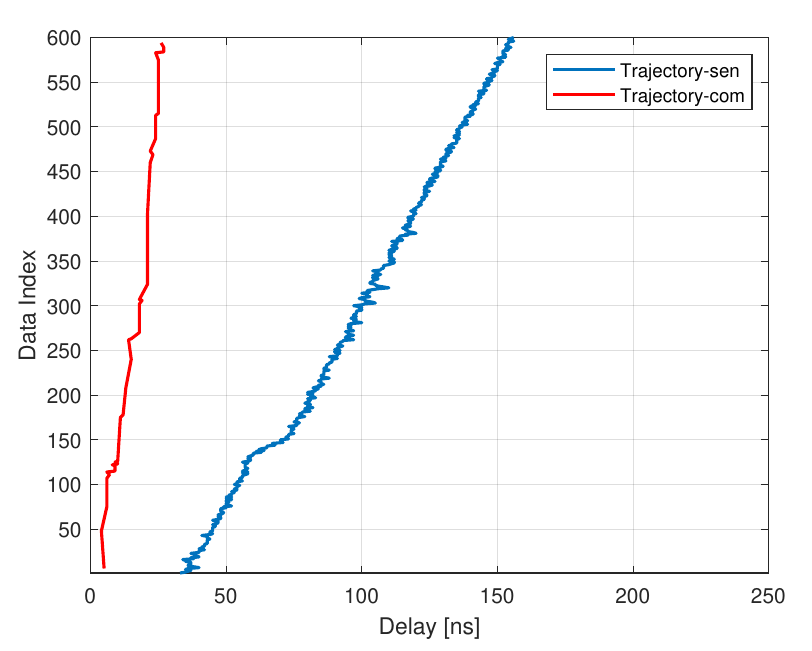}
    \caption{Trajectory of moving target multipaths in the PDP.}
    \label{fig1}
\end{figure}

\begin{figure}[t]
    \centering
        \includegraphics[width=1\linewidth]{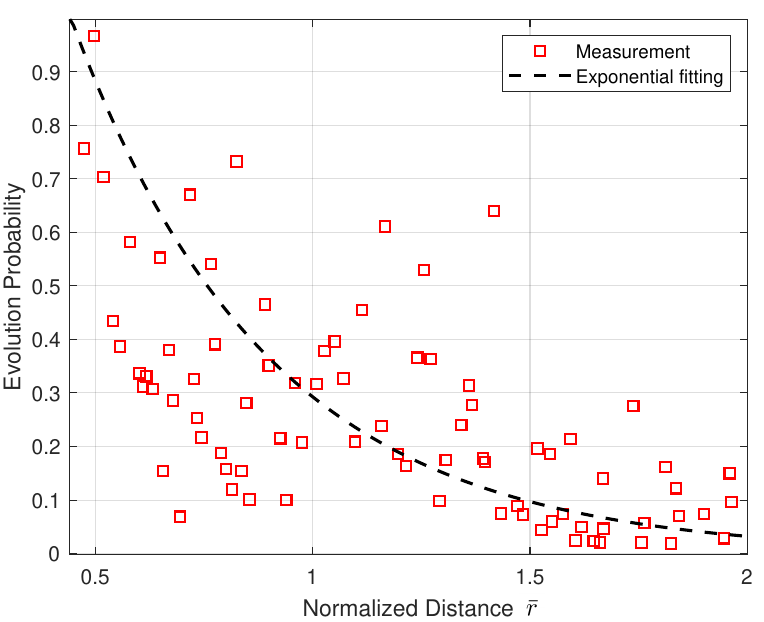}
    \caption{Evolution probability fitness in ISAC channels.}
    \label{fig1}
\end{figure}

The statistically fitted evolution probability model based on measurements is shown in Fig. 10. It can be seen that evolution probability gradually decreases as normalized distance $\bar{r}$ increases. The exponential model is found to well fit measurements and the model is shown as follows:
\begin{equation}
P_{\text{evol}}(\bar{r})=\left\{\begin{array}{c}
1, \bar{r} \leq 0.441 \\
2.664 \cdot e^{-2.208 \cdot \bar{r}}, \bar{r}>0.441
\end{array}\right.
\end{equation}
When $\bar{r}$ is less than 0.441, evolution probability $P_{\text{evol}}$ is always 1, indicating that all clusters in communication channel can be considered as shared sensing clusters. When $\bar{r}$ is larger than 0.441, evolution probability follows the exponential fitting. According to statistical model, shared sensing clusters can be generated based on the existing clusters at different spatial location in communication channels, which can be integrated into an ISAC channel with correlation properties.

\subsection{Newborn Sensing Cluster Distribution}

\begin{figure}[t]
\centering
\subfigure[]{\includegraphics[width=1.7in]{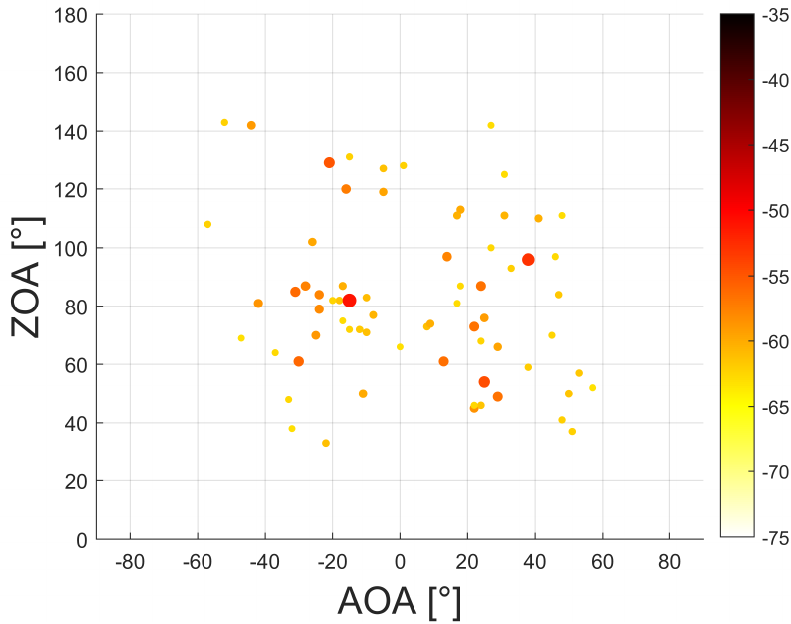}}
\subfigure[]{\includegraphics[width=1.7in]{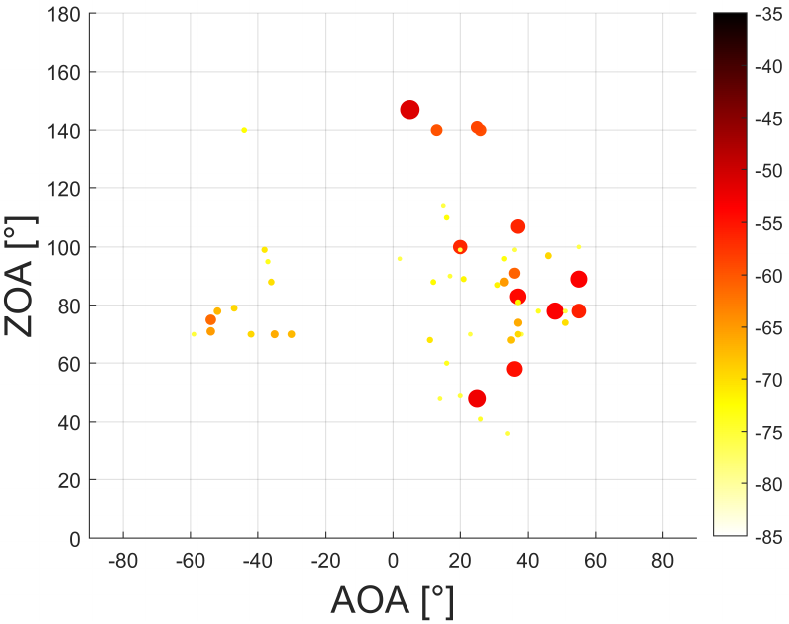}}
\subfigure[]{\includegraphics[width=3.5in]{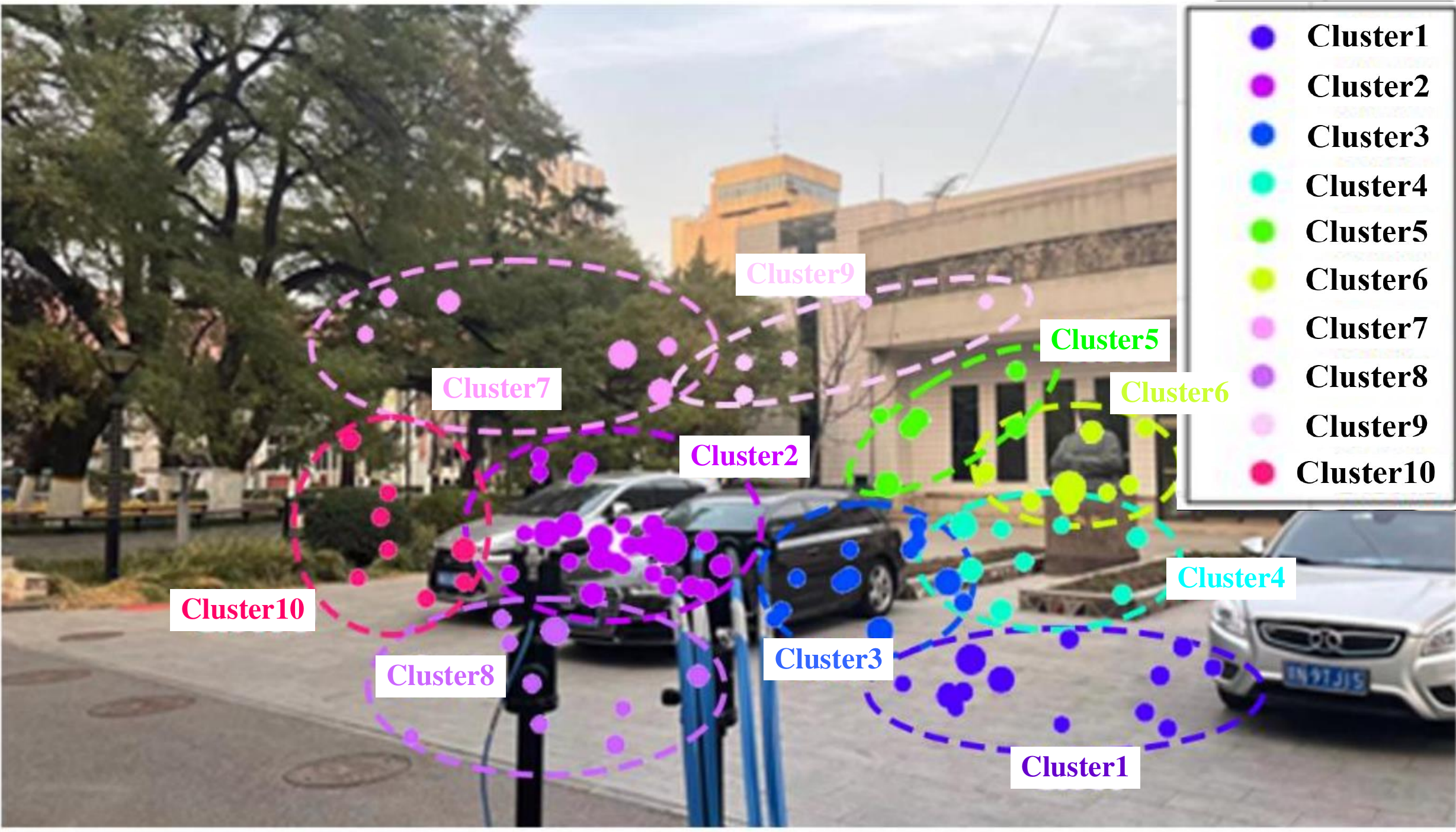}}
\subfigure[]{\includegraphics[width=3.5in]{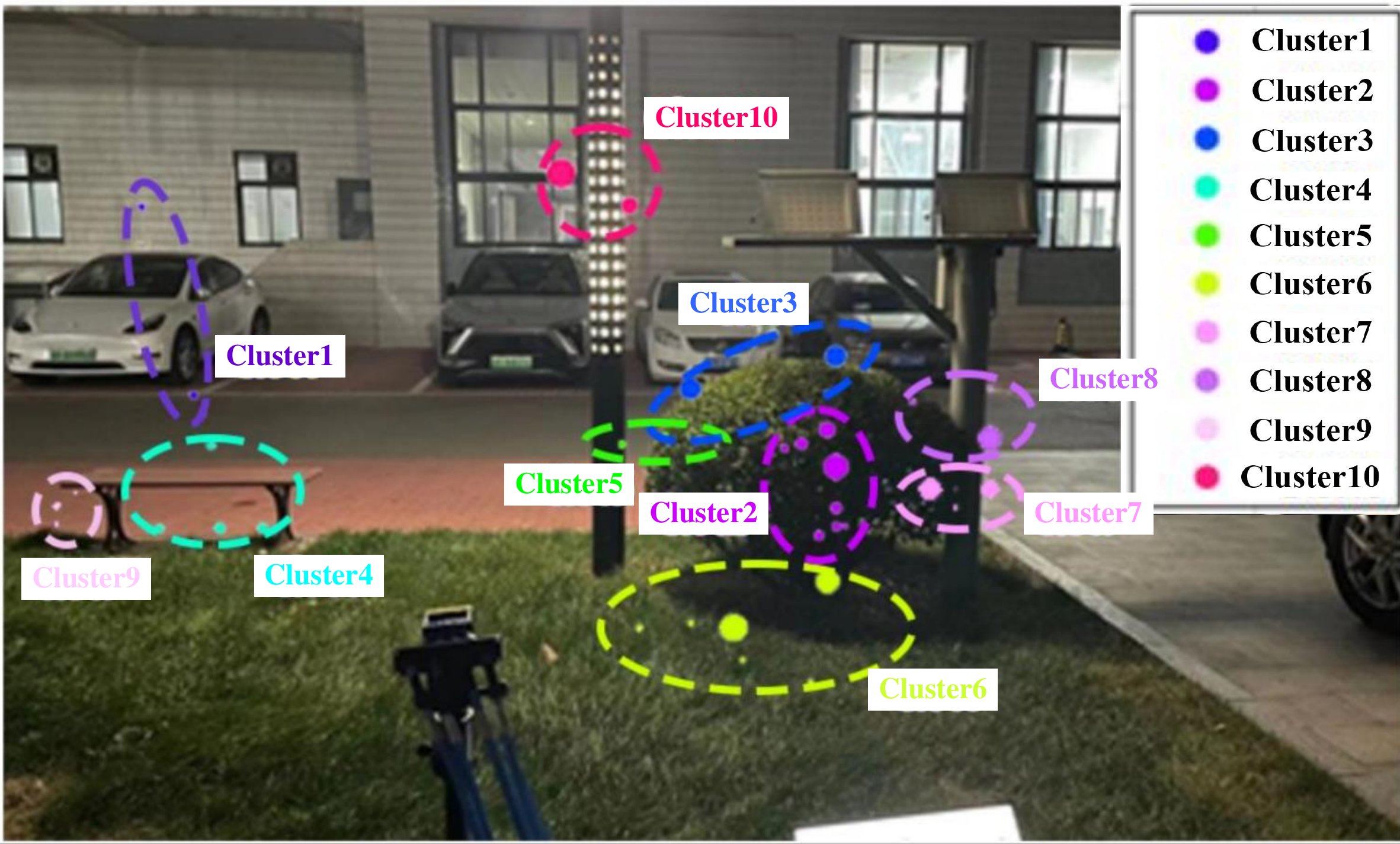}}
\caption{Estimated MPCs in static measurement scenario. (a) Estimated MPCs in sensing channel. (b) Estimated MPCs in communication channel. (c) Clustering results and environmental mapping in sensing channel, corresponding to (a). (d) Clustering results and environmental mapping in communication channel, corresponding to (b).}
\end{figure}

\begin{figure}[!htp]
\centering
\subfigure[]{\includegraphics[width=1.7in]{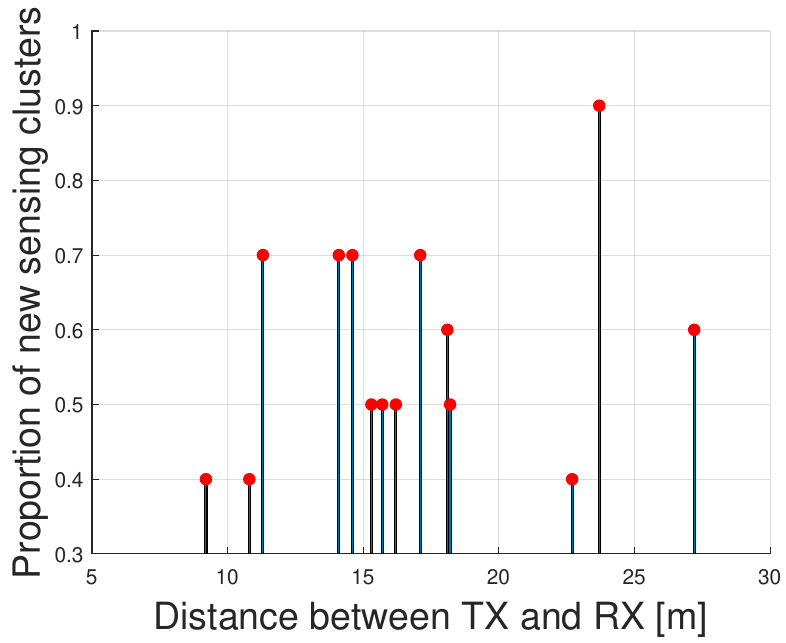}}
\subfigure[]{\includegraphics[width=1.7in]{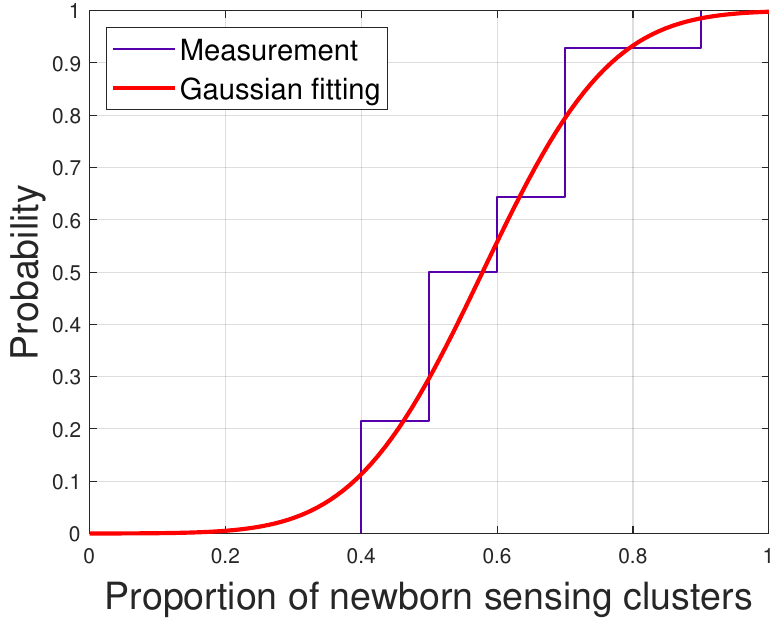}}
\caption{Distribution of newborn sensing clusters in ISAC channels. (a) Proportion variation with distance between TX and RX. (b)  CDF of proportion of newborn sensing clusters, with the normal distribution fitting result.}
\end{figure}

In order to generate newborn sensing clusters in ISAC channels, multipath estimation and environmental mapping in static scenario are provided, as shown in Fig. 11. The SAGE algorithm is employed to jointly estimate multipath parameters of communication and sensing channels. The K-Power-Means clustering algorithm \cite{Ref45,Ref46} is then employed to cluster multipaths according to delay, amplitude, AOA and ZOA. For each measurement point, number of clusters is set to 10 for comparison. Considering comparability of clustering process, the clustering results in Fig. 11 are conducted in the same environment, where the clustering results of sensing channel and communication channel are mapped to the same group of scatterers in the environment. All clustering results have been verified to ensure accuracy consistency with the actual environment. Based on multipath estimation of AOA, ZOA, power and delay, combined with visual information recorded in measurements, each cluster can be matched with a corresponding physical scatterer. The same processing flow is applied to both sensing and communication channels, and distribution of newborn sensing clusters for each measurement point is statistically analyzed.

Fig. 11(a) and Fig. 11(c) show the estimated MPCs and environmental mapping of sensing channels, where the corresponding scatterers mainly include vehicles (cluster 2 and 3), buildings (cluster 5 and 9), trees (cluster 7), ground (cluster 1 and 8), statues (cluster 4 and 6), and bushes (cluster 10). Fig. 11(b) and Fig. 11(d) show the estimated MPCs and environmental mapping of communication channels on measurement point 8, where the corresponding scatterers mainly include vehicles (cluster 1), bushes (cluster 2, 3, 5, 7 and 8), street lamp (cluster 10), bench (cluster 4 and 9), and lawns (cluster 6). Due to space limitations, the estimated MPCs and environmental mapping on other measurement points are not presented. Based on the mapping results for physical scatterer, distribution of newborn sensing clusters in ISAC channels is given as in Fig. 12. Furthermore, Fig. 12(a) shows proportion of newborn sensing clusters on different measurement points. It can be seen that number of newborn sensing clusters has a relatively weak correlation with distance. Therefore, distribution of newborn sensing clusters is modeled as independent of distance. Fig. 12(b) shows cumulative distribution function (CDF) of proportion of newborn sensing clusters, which is well fitted by the truncated Gaussian distribution with a mean of 0.578, a variance of 0.021, and a truncation range of [0,1].

\begin{figure}[t]
    \centering
        \includegraphics[width=1\linewidth]{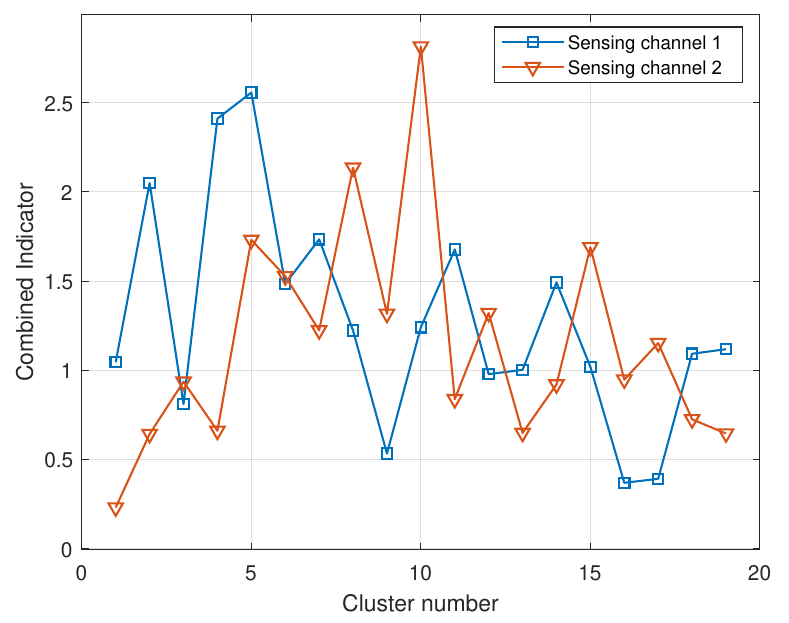}
    \caption{Clustering performance with combined indicator for different numbers.}
    \label{fig1}
\end{figure}

\subsection{Sensing Cluster Global Number}

\begin{figure*}[t]
\centering
\subfigure[]{\includegraphics[width=2.3in]{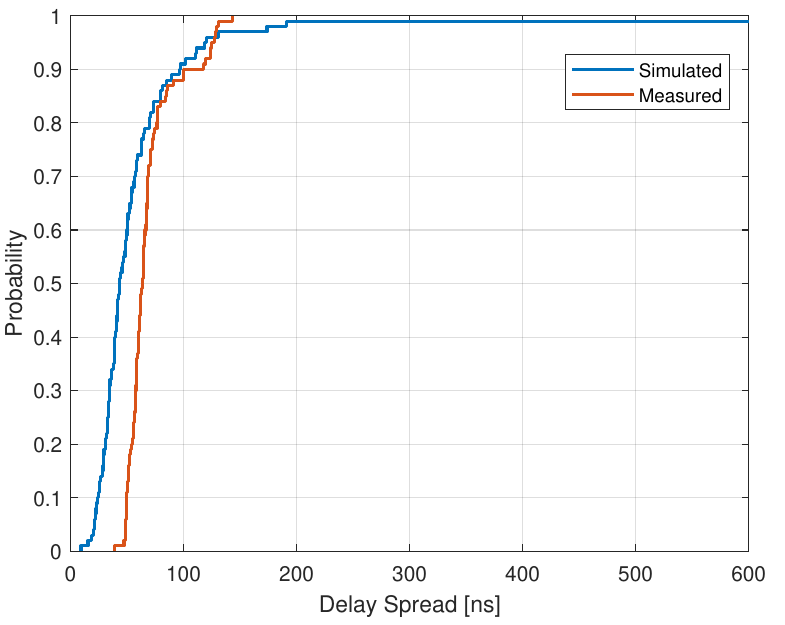}}
\subfigure[]{\includegraphics[width=2.3in]{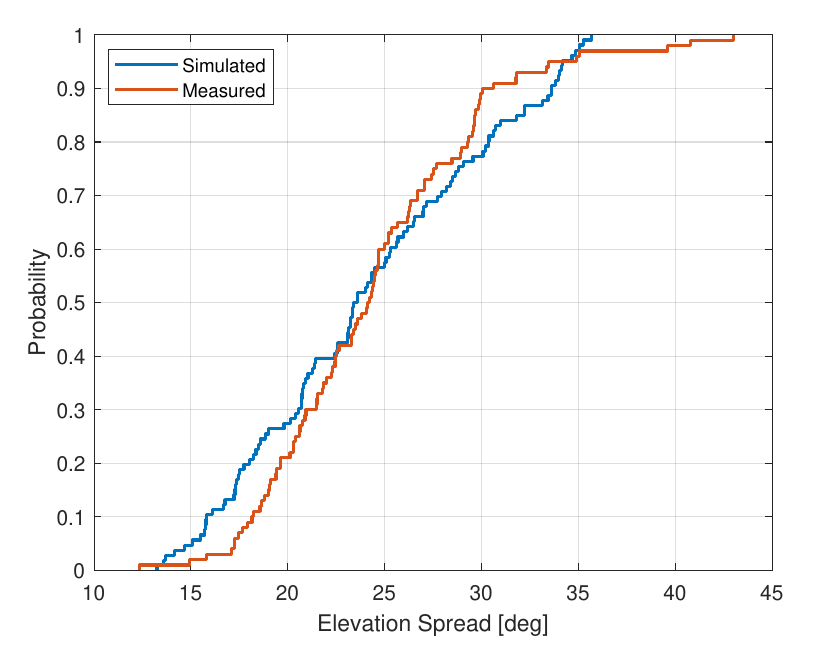}}
\subfigure[]{\includegraphics[width=2.3in]{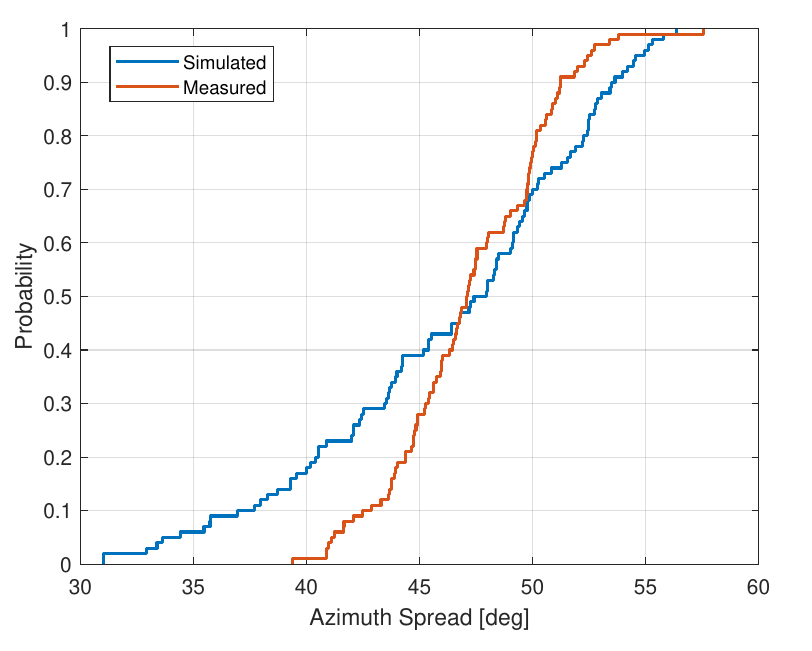}}
\caption{Model comparisons between simulated and measured channels. (a) RMS delay spread. (b) RMS elevation spread. (c) RMS azimuth spread.}
\end{figure*}

Sensing cluster is limited on global number and it is necessary to merge redundant sensing clusters. In order to determine global number in sensing channels, which is expected to be compatible to 3GPP standard channel, this subsection focus on relationship between communication and sensing channels in term of cluster numbers.

In order to obtain the optimal number of clusters, the Calinski Harabasz index (CH) and Davies Bouldin criterion (DB) are used to evaluate clustering performance\cite{Ref47, Ref48}. CH index represents ratio of inter cluster distance to intra cluster distance, calculated as follows:
\begin{equation}
C H=\frac{\operatorname{tr}\left(B_k\right)(N-K)}{\operatorname{tr}\left(W_k\right)(K-1)}
\end{equation}
where $B_k$ is covariance matrix between cluster, and $W_k$ is covariance matrix within cluster. They are calculated as follows:
\begin{equation}
B_k=\sum_{q=1}^k n_q\left(c_q-c_e\right)\left(c_q-c_e\right)^T
\end{equation}
\begin{equation}
W_k=\sum_{q=1}^k \sum_{x \in C_q}\left(x-c_q\right)\left(x-c_q\right)^T
\end{equation}
where $c_q$ represents center point of cluster $q$, $c_e$ represents center point of dataset, $n_q$ represents number of multipaths in cluster $q$, and $C_q$ represents dataset of cluster $q$. Generally speaking, a larger CH indicates better clustering performance.

DB index is calculated by dividing average distance within any two clusters by the distance between clusters, and taking the maximum value. The calculation is as follows:
\begin{equation}
D B=\frac{1}{k} \sum_{i=1}^k \max _{i \neq j, i, j \in[1, K]} \frac{s_i+s_j}{M_{i j}}
\end{equation}
where $s_i$ represents dispersion of each multipath in cluster, $M_{ij}$ represents center distance between cluster $i$ and cluster $j$, as follows:
\begin{equation}
s_i=\left\{\frac{1}{n} \sum_{j=1}^n\left|X_{i j}-A_i\right|^q\right\}^{\frac{1}{q}}
\end{equation}
\begin{equation}
M_{i j}=\left\{\sum_{k=1}^K\left|a_{k i}-a_{k j}\right|^q\right\}^{\frac{1}{q}}
\end{equation}
where $X_{ij}$ represents the $j$-th multipath component in the $i$-th cluster, $A_i$ represents center of the $i$-th cluster, $n$ represents number of multipaths in the $i$-th cluster, $a_{ki}$ represents the $k$-th dimensional value of center of the $i$-th cluster, $q = 1$ is used to calculate mean distance from each point to center. In general, a smaller DB indicates better clustering performance.

The two indicators are jointly used to obtain the optimal number of clusters, and a combined indicator (CI) index, i.e., a trade-off method for DB and CH indicators, is used, as follows:
\begin{equation}
K=\arg \max _K\left\{\left[\frac{\overline{DB}}{D B(K)}+\frac{C H(K)}{\overline{C H}}\right]\right\}
\end{equation}
where $\overline{DB}$ and $\overline{C H}$ are mean values of indicators. As shown in Fig. 13, number range of clusters in sensing channel is set as [2:20]. For sensing channel with different directions, the optimal number of clusters are 5 and 10, respectively. The average value of them is 7.5. Correspondingly, the optimal number of clusters in communication channel is [2, 4, 5, 5, 5, 5, 4, 6, 9, 4, 11, 2, 2, 16], with an average of 5.74. The ratio of number of sensing clusters and communication clusters is 1.32. Therefore, according to 3GPP TR38.901, in rural macro (RMa) scenario, number of clusters remains 11 and 10; in urban micro (UMi) scenario, number of clusters remains 12  and 19; and in urban macro (UMa) scenario, number of clusters remains 12 and 20. Besides, in all scenarios, there are 20 multipaths within each cluster. The corresponding global number of sensing cluster in ISAC channels is shown in Table III, including UMi, UMa, and RMa scenarios, respectively. Furthermore, with different simulation parameters, number of clusters in the sensing channel will thus not significantly change.

\begin{table}[]
\centering
\caption{Global number of sensing cluster.} \label{Table1Label}
\scalebox{1}{
\begin{tabular}{cccc}
\hline
Scenario             & Propagation & \begin{tabular}[c]{@{}c@{}}Communication\\ (as 3GPP TR38.901)\end{tabular} & Sensing \\ \hline
\multirow{2}{*}{UMi} & LOS         & 12                                                                               & 16      \\
                     & NLOS        & 19                                                                               & 26      \\ 
\multirow{2}{*}{UMa} & LOS         & 12                                                                               & 16      \\
                     & NLOS        & 20                                                                               & 27      \\ 
\multirow{2}{*}{RMa} & LOS         & 11                                                                               & 15      \\
                     & NLOS        & 10                                                                               & 14      \\ \hline
\end{tabular}}
\end{table}

\begin{table*}[]
\centering
\caption{Detailed simulation parameters.}
\scalebox{1.1}{
\begin{tabular}{|c|c|ccc|}
\hline
Simulation parameters & Measurement validation & \multicolumn{3}{c|}{Multi-link simulation}                                                                                                                                                                                                                                                                                                                      \\ \hline
Scenario              & UMi                    & \multicolumn{1}{c|}{UMi}                                                                                                   & \multicolumn{1}{c|}{UMa}                                                                                                   & RMa                                                                                                   \\ \hline
Frequency             & 28G                    & \multicolumn{1}{c|}{28G}                                                                                                   & \multicolumn{1}{c|}{28G}                                                                                                   & 28G                                                                                                   \\ \hline
Bandwidth             & 1G                     & \multicolumn{1}{c|}{1G}                                                                                                    & \multicolumn{1}{c|}{1G}                                                                                                    & 1G                                                                                                    \\ \hline
TX/SX heights         & 5m                     & \multicolumn{1}{c|}{\begin{tabular}[c]{@{}c@{}}BS1-20m, \\ BS2-35m\end{tabular}}                                           & \multicolumn{1}{c|}{\begin{tabular}[c]{@{}c@{}}BS1-20m, \\ BS2-35m\end{tabular}}                                           & \begin{tabular}[c]{@{}c@{}}BS1-20m, \\ BS2-35m\end{tabular}                                           \\ \hline
RX heights            & 1.5m                   & \multicolumn{1}{c|}{\begin{tabular}[c]{@{}c@{}}UT1-1.5m, \\ UT2-3.5m, \\ UT3-1m\end{tabular}}                              & \multicolumn{1}{c|}{\begin{tabular}[c]{@{}c@{}}UT1-1.5m, \\ UT2-3.5m, \\ UT3-1m\end{tabular}}                              & \begin{tabular}[c]{@{}c@{}}UT1-1.5m, \\ UT2-3.5m, \\ UT3-1m\end{tabular}                              \\ \hline
TX/SX coordinate      & {[}0,0{]}m             & \multicolumn{1}{c|}{\begin{tabular}[c]{@{}c@{}}BS1-{[}100,100{]}m, \\ BS2-{[}150,150{]}m\end{tabular}}                     & \multicolumn{1}{c|}{\begin{tabular}[c]{@{}c@{}}BS1-{[}100,100{]}m, \\ BS2-{[}150,150{]}m\end{tabular}}                     & \begin{tabular}[c]{@{}c@{}}BS1-{[}100,100{]}m, \\ BS2-{[}150,150{]}m\end{tabular}                     \\ \hline
RX coordinate         & {[}8,8{]}m             & \multicolumn{1}{c|}{\begin{tabular}[c]{@{}c@{}}UT1-{[}50,50{]}m, \\ UT2-{[}20,180{]}m, \\ UT3-{[}170,30{]}m,\end{tabular}} & \multicolumn{1}{c|}{\begin{tabular}[c]{@{}c@{}}UT1-{[}50,50{]}m, \\ UT2-{[}20,180{]}m, \\ UT3-{[}170,30{]}m,\end{tabular}} & \begin{tabular}[c]{@{}c@{}}UT1-{[}50,50{]}m, \\ UT2-{[}20,180{]}m, \\ UT3-{[}170,30{]}m,\end{tabular} \\ \hline
\end{tabular}}
\end{table*}

\section{Channel Simulation and validation}

Multi-link simulation and measurement validation are presented in this section. To facilitate comparison with measurements, we set antenna height, positions, and other simulation parameters to be the same as those in the measurements for validation. In the case of multi-link simulation, a general situation is considered, and the simulation parameters are not required to be the same as in measurements. The detailed simulation parameters are shown in Table IV.

\subsection{Measurement Validation}

In this subsection, three statistics, i.e., root-mean square (RMS) delay spread, elevation spread and azimuth spread of arrival, are used to validate the proposed model. In order to compare with measurements, we set antenna height, positions and other simulation parameters to be the same as in measurements.

In Fig. 14, CDF comparisons of delay spread, elevation spread and azimuth spread between simulated and measured channels are presented. Due to the fact that positions of scatterers in simulation are randomly generated, although UE parameters are set as consistent as possible with measurements, the generated channel parameters, especially spatial distribution of scatterers, cannot fully match with measurements. However, the agreement is fairly good and reasonable. The errors between simulated and measured channels with 90$\%$ CDF probability are 2.74$\%$, 9.84$\%$ and 4.68$\%$ for delay spread, elevation spread and azimuth spread respectively. Overall, the simulated channel based on the proposed ISAC channel models can reflect characteristics of actual channel well.

\subsection{Multi-Link Simulation}
\begin{figure}[t]
    \centering
        \includegraphics[width=1\linewidth]{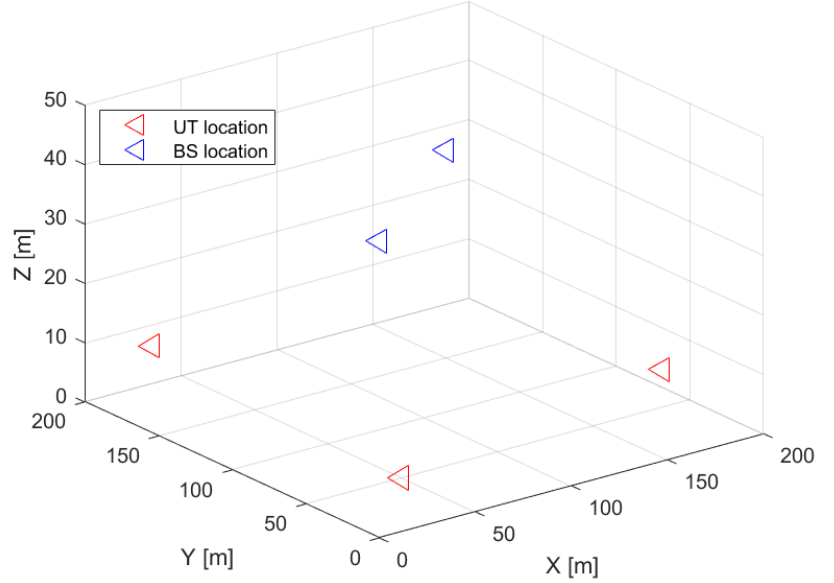}
    \caption{Layout of BS and UT in multi-link simulation of ISAC scenario}
    \label{fig1}
\end{figure}

\begin{figure*}[t]
\centering
\subfigure[]{\includegraphics[width=2.35in]{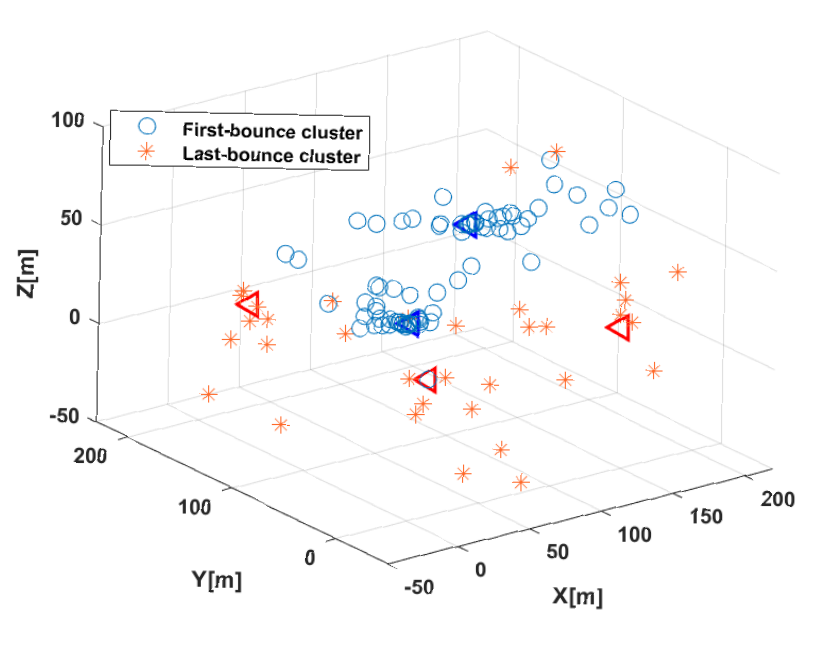}}
\subfigure[]{\includegraphics[width=2.35in]{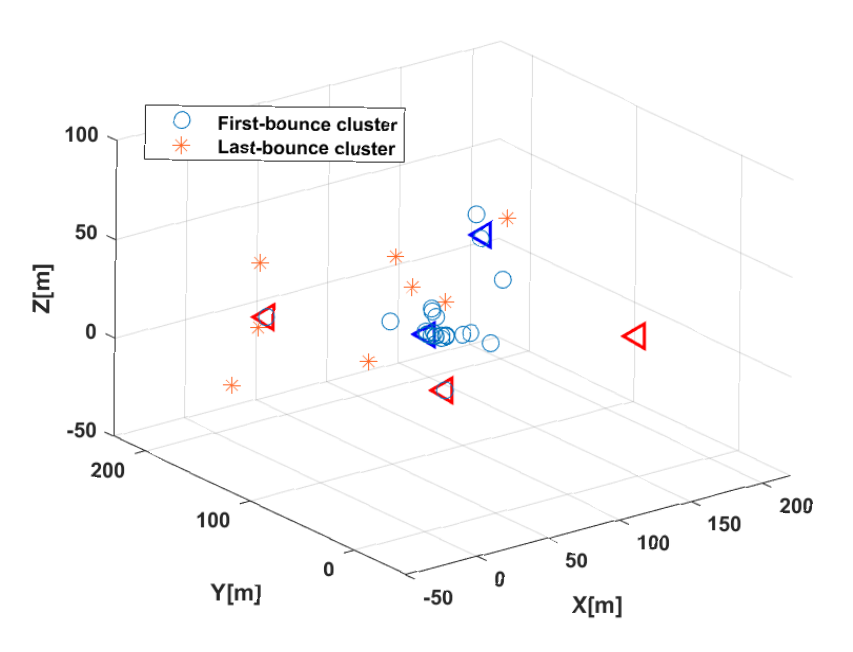}}
\subfigure[]{\includegraphics[width=2.35in]{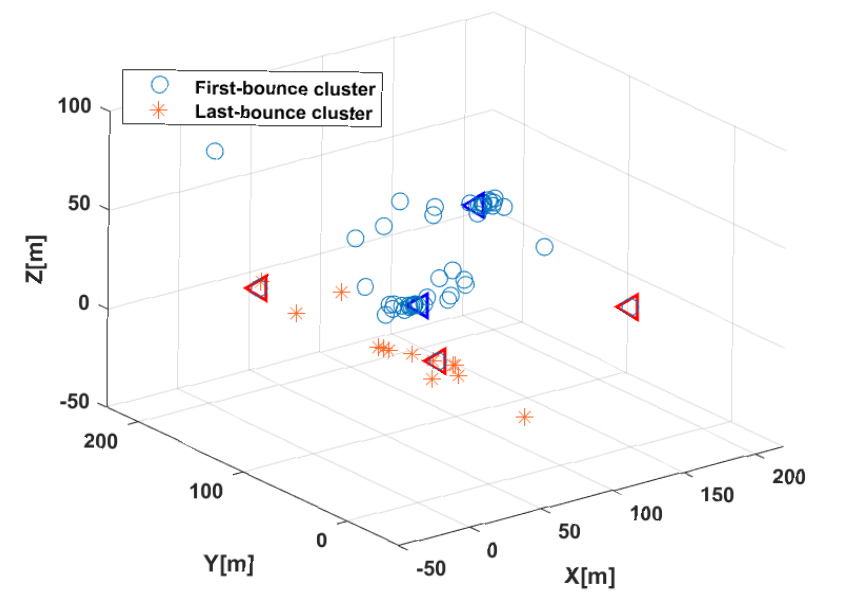}}
\caption{Communication cluster distribution of ISAC channel models. (a) UMi scenario. (b) UMa scenario. (c) RMa scenario.}
\end{figure*}

\begin{figure*}[t]
\centering
\subfigure[]{\includegraphics[width=2.35in]{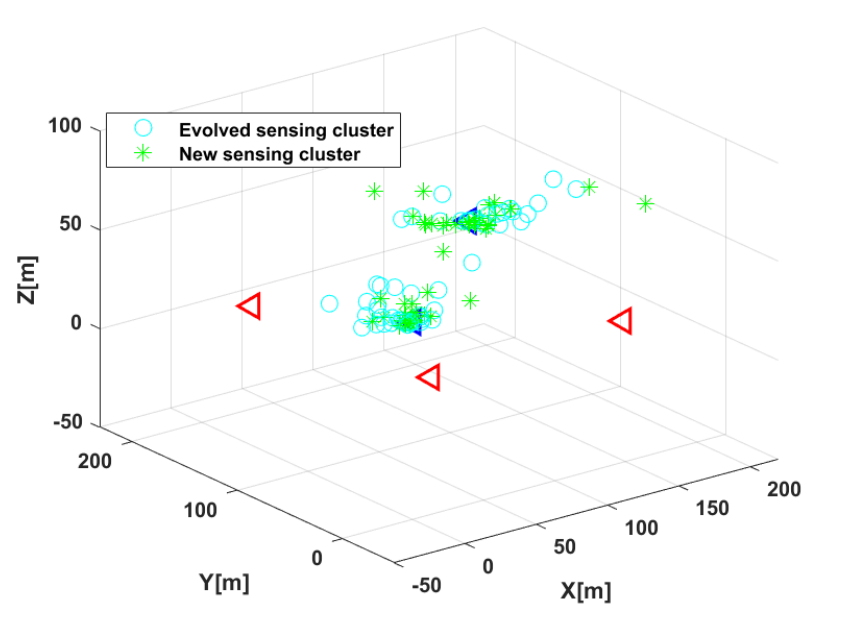}}
\subfigure[]{\includegraphics[width=2.35in]{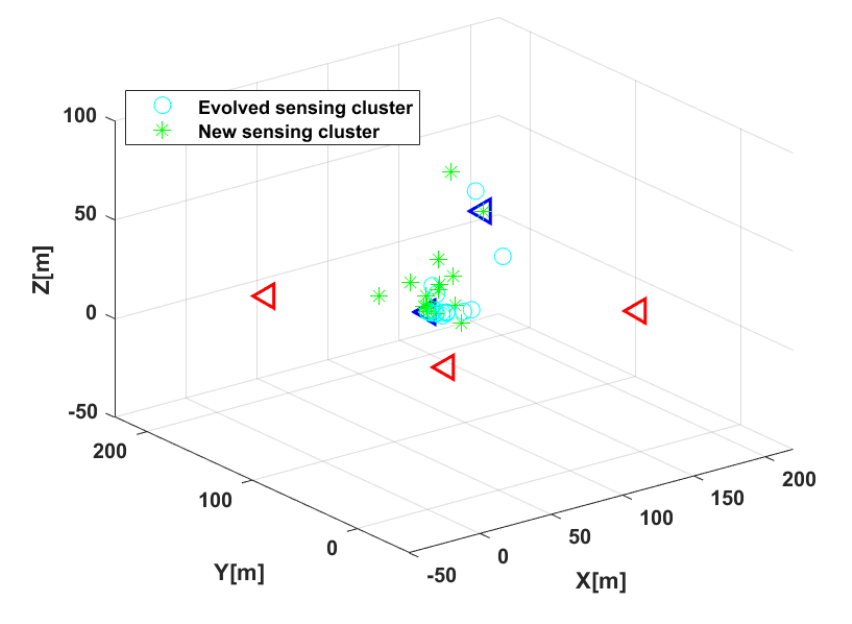}}
\subfigure[]{\includegraphics[width=2.35in]{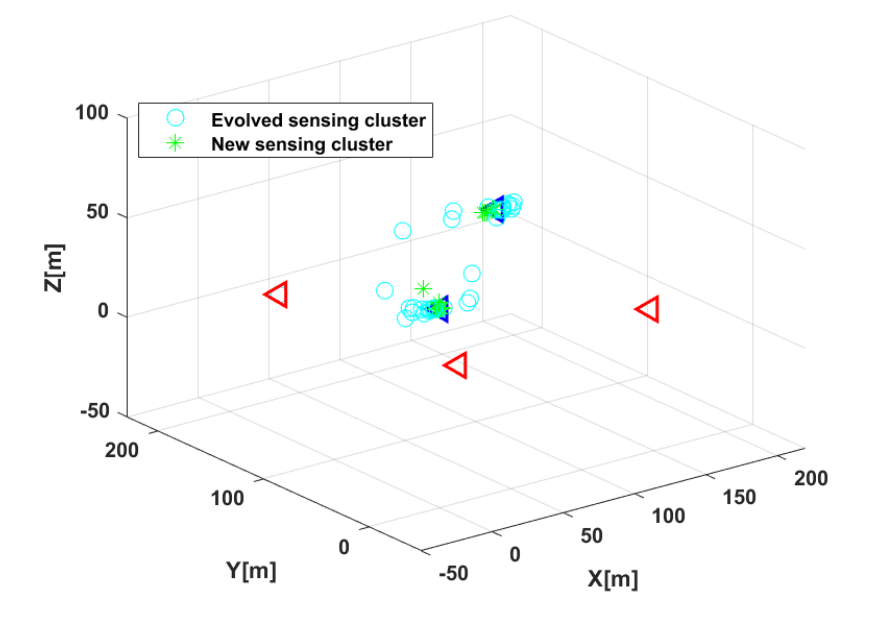}}
\caption{Sensing cluster distribution of ISAC channel models. (a) UMi scenario. (b) UMa scenario. (c) RMa scenario.}
\end{figure*}

\begin{figure*}[t]
\centering
\subfigure[]{\includegraphics[width=2.3in]{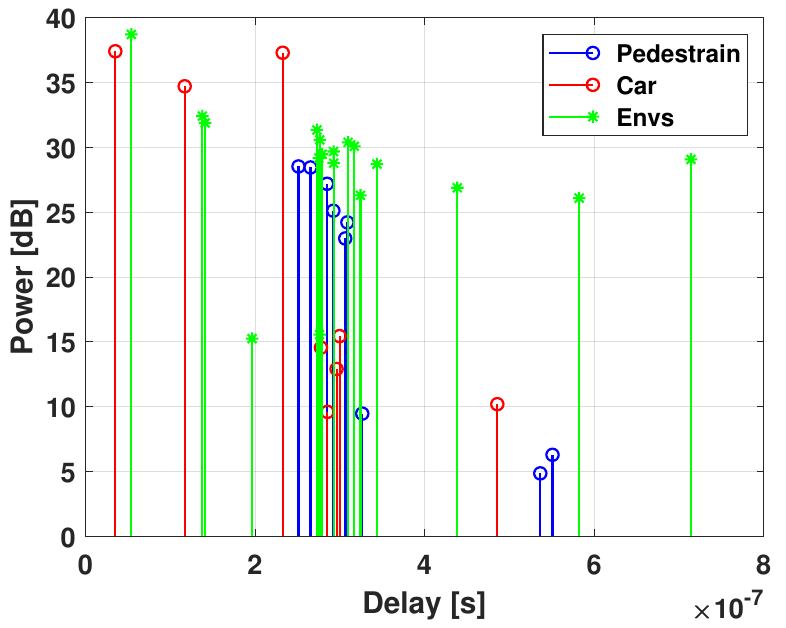}}
\subfigure[]{\includegraphics[width=2.3in]{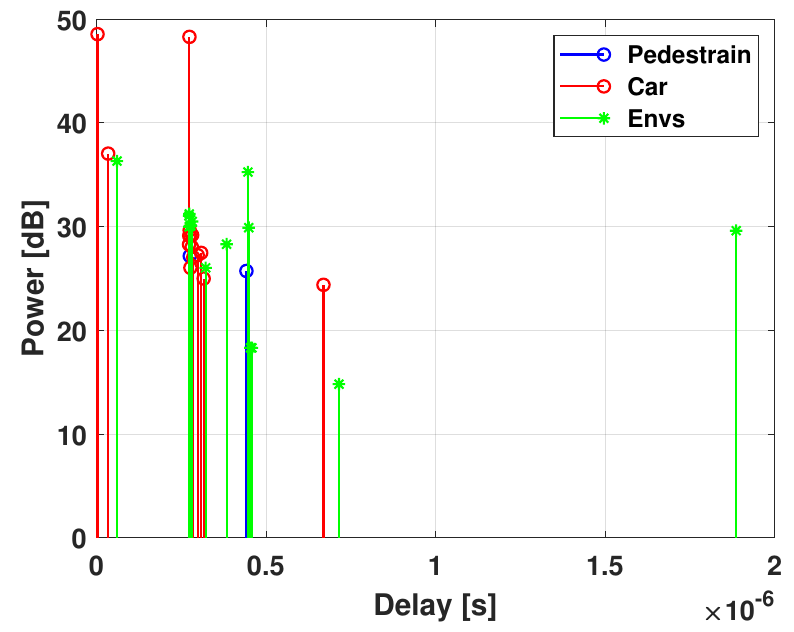}}
\subfigure[]{\includegraphics[width=2.3in]{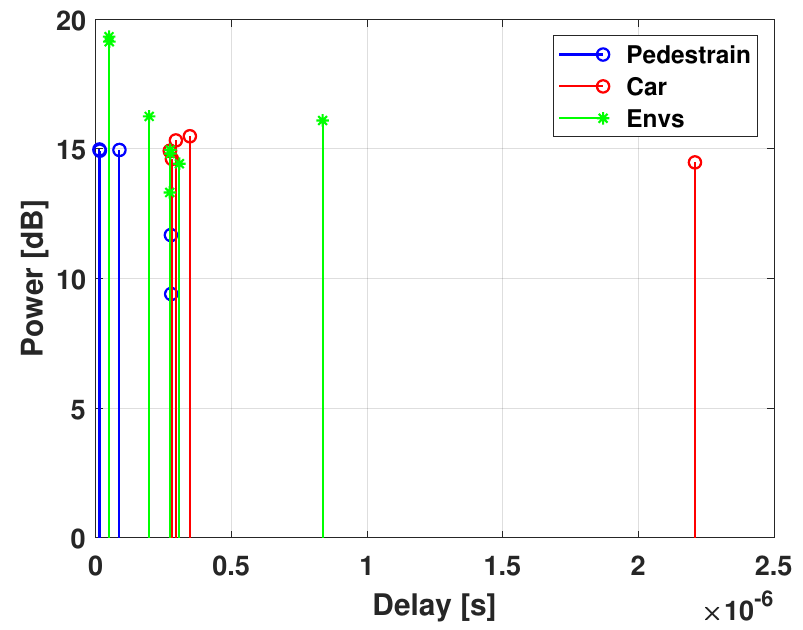}}
\caption{Generated MPCs based on the ISAC channel models. (a) UMi scenario. (b) UMa scenario. (c) RMa scenario.}
\end{figure*}

This subsection presents simulation results of multi-link layout in ISAC scenario, including distribution of communication and sensing clusters and MPCs. Fig. 15 shows initial layout and there are two BSs and three UTs with BS1 located at [100, 100, 20] m, BS2 located at [150, 150, 35] m, UT1 located at [50, 50, 1.5] m, UT2 located at [20, 180, 3.5] m, and UT3 located at [170, 30, 1] m. Carrier frequency is 28 GHz, bandwidth is 1 GHz, and simulation scenarios are UMi, UMa, and RMa scenarios, respectively.

RCS values of similar objects vary significantly due to specific conditions, such as humans worn with different clothing and cars with different materials. According to existing researches on RCS measurements for different objects, the uniform distribution with special ranges is used to generate RCS values for a single scatterer. The surrounding scatterers in simulation are given as follows: 30$\%$ for vehicles, 20$\%$ for pedestrians, and 50$\%$ for other environmental objects (buildings, trees, etc.). RCS values of pedestrians, vehicles and environmental objects in ISAC channels are modeled as uniform distribution of [-20, 0] decibel per square mete (dBsm), [-5, 25] dBsm, and  [-50, 50] dBsm, respectively\cite{Ref49,Ref50,Ref51}. For each ray in sensing channels, RCS is applied independently.

Fig. 16 shows distribution of communication clusters in UMi, UMa, and RMa scenarios, respectively, where different colors represent different communication links. Fig. 17 shows the corresponding distribution of sensing clusters in UMi, UMa, and RMa scenarios, respectively, where two colors represent shared sensing clusters and newborn sensing clusters. It can be seen that, ISAC channels represent spatial characteristics well and clusters perceived by one BS are not limited in own communication links. For distant communication clusters, they are more difficult to be considered as shared sensing clusters in terms of evolution probability. Due to limitations in ISAC channel measurements, we refer to 3GPP channel models to simplify our modeling procedure. Specifically, in the ISAC channel modeling procedure, the communication channel is first generated using the 3GPP model procedure. Various simulation parameters, such as elevation, altitude, and direction, influence the spatial distribution of communication clusters. Consequently, the distribution of sensing clusters is also affected based on correlatively modeling. As a result, the arrival and departure angle distributions in the ISAC channel model are altered. Fig. 18 shows the corresponding MPCs of sensing channels in UMi, UMa, and RMa scenarios, respectively, where different colors represent multipaths from different scatterers. It can be seen that MPCs provided in the proposed models can well perform sensing characteristics dependant on physical properties.

\section{Conclusion}
In this paper, a cluster-based statistical channel model is proposed for ISAC scenarios, and it considers distributions of clusters in both communication and sensing channels. Based on procedure of 3GPP standard channel model, we present an extension to incorporate sensing channels into the existing communication channel modeling framework. Measurement campaigns at 28 GHz are conducted for parameterization. Based on measurements, clustering and tracking algorithms are used to extract and analyze clusters in ISAC channels. Sensing clusters are divided into shared sensing clusters and newborn sensing clusters. Shared sensing clusters come from the existing clusters in communication channels based on evolution probability, which is modeled as exponential distribution. Newborn sensing clusters only contribute to sensing channel, which have a truncated Gaussian distribution. Furthermore, this paper presents model implementation, and validates the model by comparing simulations and measurements. The proposed model provides a simulation approach for performance evaluation of ISAC technologies.




\ifCLASSOPTIONcaptionsoff
  \newpage
\fi



\begin{thebibliography}{00}


\bibitem{Ref1}
L. U. Khan, W. Saad, D. Niyato, Z. Han and C. S. Hong, ``Digital-Twin-Enabled 6G: Vision, Architectural Trends, and Future Directions," {\it{IEEE Communications Magazine}}, vol. 60, no. 1, pp. 74-80, 2022.

\bibitem{Ref2}
Z. Wei, F. Liu, C. Masouros, N. Su and A. P. Petropulu, ``Toward Multi-Functional 6G Wireless Networks: Integrating Sensing, Communication, and Security," {\it{IEEE Communications Magazine}}, vol. 60, no. 4, pp. 65-71, 2022.

\bibitem{Ref3}
R. He et al., ``Propagation Channels of 5G Millimeter-Wave Vehicle-to-Vehicle Communications: Recent Advances and Future Challenges," {\it{IEEE Vehicular Technology Magazine}}, vol. 15, no. 1, pp. 16-26, 2020.

\bibitem{Ref4}
Y. Cui, F. Liu, X. Jing and J. Mu, ``Integrating Sensing and Communications for Ubiquitous IoT: Applications, Trends, and Challenges," {\it{IEEE Network}}, vol. 35, no. 5, pp. 158-167, 2021.

\bibitem{Ref5}
R. He, B. Ai, G. L. Stüber, G. Wang and Z. Zhong, ``Geometrical-Based Modeling for Millimeter-Wave MIMO Mobile-to-Mobile Channels," {\it{IEEE Transactions on Vehicular Technology}}, vol. 67, no. 4, pp. 2848-2863, 2018.

\bibitem{Ref6}
M. Yang et al., ``Measurements and Cluster-Based Modeling of Vehicle-to-Vehicle Channels With Large Vehicle Obstructions," {\it{IEEE Transactions on Wireless Communications}}, vol. 19, no. 9, pp. 5860-5874, 2020.

\bibitem{Ref7}
C. Huang et al., ``Artificial Intelligence Enabled Radio Propagation for Communications—Part II: Scenario Identification and Channel Modeling," {\it{IEEE Transactions on Antennas and Propagation}}, vol. 70, no. 6, pp. 3955-3969, 2022.

\bibitem{Ref8}
Q. Zhang, H. Sun, X. Gao, X. Wang and Z. Feng, ``Time-Division ISAC Enabled Connected Automated Vehicles Cooperation Algorithm Design and Performance Evaluation," {\it{IEEE Journal on Selected Areas in Communications}}, vol. 40, no. 7, pp. 2206-2218, 2022.

\bibitem{Ref9}
C. Liu et al., ``Learning-Based Predictive Beamforming for Integrated Sensing and Communication in Vehicular Networks," {\it{IEEE Journal on Selected Areas in Communications}}, vol. 40, no. 8, pp. 2317-2334, 2022.

\bibitem{Ref10}
Y. Xiong, F. Liu, Y. Cui, W. Yuan and T. X. Han, ``Flowing the Information from Shannon to Fisher: Towards the Fundamental Tradeoff in ISAC," {\it{GLOBECOM 2022-2022 IEEE Global Communications Conference, Rio de Janeiro, Brazil}}, 2022, pp. 5601-5606.

\bibitem{Ref11}
Z. Zhang et al., ``A General Channel Model for Integrated Sensing and Communication Scenarios," {\it{IEEE Communications Magazine}}, vol. 61, no. 5, pp. 68-74, 2023.

\bibitem{Ref12}
L. Pang, J. Zhang, Y. Zhang, X. Huang, Y. Chen and J. Li, ``Investigation and comparison of 5G channel models: From QuaDRiGa, NYUSIM, and MG5G perspectives," {\it{Chinese Journal of Electronics}}, vol. 31, no. 1, pp. 1-17, 2022.

\bibitem{Ref13}
L. Xiong, Z. Zhang and D. Yao. ``A novel real time channel prediction algorithm in high-speed scenario using convolutional neural network," {\it{Wireless Networks}}, vol. 28, pp. 621–634, 2022.

\bibitem{Ref14}
M. Yang et al., ``Dynamic V2V Channel Measurement and Modeling at Street Intersection Scenarios," {\it{IEEE Transactions on Antennas and Propagation}}, vol. 71, no. 5, pp. 4417-4432, 2023.

\bibitem{Ref15}
R. He, B. Ai, G. Wang, M. Yang, C. Huang and Z. Zhong, ``Wireless Channel Sparsity: Measurement, Analysis, and Exploitation in Estimation," {\it{ IEEE Wireless Communications}}, vol. 28, no. 4, pp. 113-119, 2021.

\bibitem{Ref16}
H. Zhang, R. He, B. Ai, S. Cui and H. Zhang, ``Measuring Sparsity of Wireless Channels," {\it{IEEE Transactions on Cognitive Communications and Networking}}, vol. 7, no. 1, pp. 133-144, 2021.


\bibitem{Ref17}
P. Zhu, X. Yin, J. Rodríguez-Pineiro, Z. Chen, P. Wang and G. Li, ``Measurement-Based Wideband Space-Time Channel Models for 77GHz Automotive Radar in Underground Parking Lots," {\it{IEEE Transactions on Intelligent Transportation Systems}}, vol. 23, no. 10, pp. 19105-19120, 2022.

\bibitem{Ref18}
S. D. Liyanaarachchi, T. Riihonen, C. B. Barneto and M. Valkama, ``Optimized Waveforms for 5G–6G Communication With Sensing: Theory, Simulations and Experiments," {\it{IEEE Transactions on Wireless Communications}}, vol. 20, no. 12, pp. 8301-8315, 2021.

\bibitem{Ref19}
Q. Tang, J. Li, L. Wang, Y. Jia and G. Cui, ``Multipath Imaging for NLOS Targets Behind an L-Shaped Corner With Single-Channel UWB Radar," {\it{IEEE Sensors Journal}}, vol. 22, no. 2, pp. 1531-1540, 2022.

\bibitem{Ref20}
J. Wang, J. Zhang, Y. Zhang, T. Jiang, L. Yu and G. Liu, ``Empirical Analysis of Sensing Channel Characteristics and Environment Effects at 28 GHz," {\it{2022 IEEE Globecom Workshops (GC Wkshps), Rio de Janeiro, Brazil}}, 2022, pp. 1323-1328.

\bibitem{Ref21}
T. Huang, N. Shlezinger, X. Xu, Y. Liu and Y. C. Eldar, ``MAJoRCom: A Dual-Function Radar Communication System Using Index Modulation," {\it{IEEE Transactions on Signal Processing}}, vol. 68, pp. 3423-3438, 2020.

\bibitem{Ref22}
P. Kumari, J. Choi, N. González-Prelcic and R. W. Heath, ``IEEE 802.11ad-Based Radar: An Approach to Joint Vehicular Communication-Radar System," {\it{IEEE Transactions on Vehicular Technology}}, vol. 67, no. 4, pp. 3012-3027, 2018.

\bibitem{Ref23}
Z. Cheng, S. Shi, Z. He and B. Liao, ``Transmit Sequence Design for Dual-Function Radar-Communication System With One-Bit DACs," {\it{IEEE Transactions on Wireless Communications}}, vol. 20, no. 9, pp. 5846-5860, 2021.

\bibitem{Ref24}
M. Temiz, E. Alsusa and M. W. Baidas, ``A Dual-Functional Massive MIMO OFDM Communication and Radar Transmitter Architecture," {\it{IEEE Transactions on Vehicular Technology}}, vol. 69, no. 12, pp. 14974-14988, 2020.

\bibitem{Ref25}
P. Kumari, A. Mezghani and R. W. Heath, ``JCR70: A Low-Complexity Millimeter-Wave Proof-of-Concept Platform for a Fully-Digital SIMO Joint Communication-Radar," {\it{IEEE Open Journal of Vehicular Technology}}, vol. 2, pp. 218-234, 2021.

\bibitem{Ref26}
F. Liu, C. Masouros, A. P. Petropulu, H. Griffiths and L. Hanzo, ``Joint Radar and Communication Design: Applications, State-of-the-Art, and the Road Ahead," {\it{IEEE Transactions on Communications}}, vol. 68, no. 6, pp. 3834-3862, 2020.

\bibitem{Ref27}
R. S. Thoma, C. Andrich, S. J. Myint, C. Schneider, and G. Sommerkorn, ``Characterization of multi-link propagation and bistatic target reflectivity for distributed ISAC,” 2022. [Online]. Available: https://arxiv.org/abs/2210.11840.

\bibitem{Ref28}
A. Graff, A. Ali and N. González-Prelcic, ``Measuring radar and communication congruence at millimeter wave frequencies," {\it{2019 53rd Asilomar Conference on Signals, Systems, and Computers, Pacific Grove, CA, USA}}, 2019, pp. 925-929.

\bibitem{Ref29}
Yameng Liu, Jianhua Zhang, Yuxiang Zhang, Zhiqiang Yuan, Guangyi Liu, ``A Shared Cluster-based Stochastic Channel Model for Joint Communication and Sensing Systems," 2022. [Online]. Available: https://doi.org/10.48550/arXiv.2211.06615.

\bibitem{Ref30}
C. Baquero Barneto et al., ``Millimeter-Wave Mobile Sensing and Environment Mapping: Models, Algorithms and Validation," {\it{IEEE Transactions on Vehicular Technology}}, vol. 71, no. 4, pp. 3900-3916, 2022.

\bibitem{Ref31}
A. Ali, N. González-Prelcic and A. Ghosh, ``Passive Radar at the Roadside Unit to Configure Millimeter Wave Vehicle-to-Infrastructure Links," {\it{IEEE Transactions on Vehicular Technology}}, vol. 69, no. 12, pp. 14903-14917, 2020.

\bibitem{Ref32}
C. Jiao, Z. Zhang, C. Zhong and Z. Feng, ``An Indoor mmWave Joint Radar and Communication System with Active Channel Perception," {\it{2018 IEEE International Conference on Communications (ICC), Kansas City, MO, USA}}, 2018, pp. 1-6.

\bibitem{Ref33}
G. Duggal, S. Vishwakarma, K. V. Mishra and S. S. Ram, ``Doppler-Resilient 802.11ad-Based Ultrashort Range Automotive Joint Radar-Communications System," {\it{IEEE Transactions on Aerospace and Electronic Systems}}, vol. 56, no. 5, pp. 4035-4048, 2020.

\bibitem{Ref34}
3GPP TR 38.901 version 16.1.0 Release 16, ``5G; Study on channel model for frequencies from 0.5 to 100 GHz,” Nov. 2020.

\bibitem{Ref35}
C. Huang, A. F. Molisch, R. He, R. Wang, P. Tang and Z. Zhong, ``Machine-Learning-Based Data Processing Techniques for Vehicle-to-Vehicle Channel Modeling," {\it{IEEE Communications Magazine}}, vol. 57, no. 11, pp. 109-115, 2019.

\bibitem{Ref36}
R. He et al., ``Clustering Enabled Wireless Channel Modeling Using Big Data Algorithms," {\it{IEEE Communications Magazine}}, vol. 56, no. 5, pp. 177-183, 2018.

\bibitem{Ref37}
M. Yang et al., ``A Cluster-Based Three-Dimensional Channel Model for Vehicle-to-Vehicle Communications," {\it{IEEE Transactions on Vehicular Technology}}, vol. 68, no. 6, pp. 5208-5220, 2019.

\bibitem{Ref38}
S. Jaeckel, L. Raschkowski, K. Börner and L. Thiele, ``QuaDRiGa: A 3-D Multi-Cell Channel Model With Time Evolution for Enabling Virtual Field Trials," {\it{IEEE Transactions on Antennas and Propagation}}, vol. 62, no. 6, pp. 3242-3256, 2014.

\bibitem{Ref39}
Z. Zhang et al., ``A Shared Multipath Components Evolution Model for Integrated Sensing and Communication Channels,” {\it{IEEE Antennas and Wireless Propagation Letters}}, vol. 22, no. 12, pp. 2975-2978, 2023.

\bibitem{Ref40}
Murtagh, F. and Contreras, P. ``Algorithms for hierarchical clustering: an overview," {\it{WIREs Data Mining Knowl Discov}}, vol. 2, no. 1, pp. 86-97, 2012.

\bibitem{Ref41}
Z. Zhang, R. He, M. Yang and et al., ``Millimeter Wave Channel Measurements and Analysis for Integrated Sensing and Communication Scenario," {\it{2023 IEEE International Symposium on Antennas and Propagation and USNC-URSI Radio Science Meeting (AP-S/URSI), Portland, OR, USA}}, 2023, accepted.

\bibitem{Ref42}
Y. Ji, W. Fan, P. Kyösti, J. Li and G. F. Pedersen, ``Antenna Correlation Under Geometry-Based Stochastic Channel Models," {\it{IEEE Antennas and Wireless Propagation Letters}}, vol. 18, no. 12, pp. 2567-2571, 2019.

\bibitem{Ref43}
X. Chen et al., ``Simultaneous Decoupling and Decorrelation Scheme of MIMO Arrays," {\it{IEEE Transactions on Vehicular Technology}}, vol. 71, no. 2, pp. 2164-2169, 2022.

\bibitem{Ref44}
B. H. Fleury, M. Tschudin, R. Heddergott, D. Dahlhaus and K. Ingeman Pedersen, ``Channel parameter estimation in mobile radio environments using the SAGE algorithm," {\it{IEEE Journal on Selected Areas in Communications}}, vol. 17, no. 3, pp. 434-450, 1999.

\bibitem{Ref45}
N. Czink, P. Cera, J. Salo, E. Bonek, J. -p. Nuutinen and J. Ylitalo, ``A Framework for Automatic Clustering of Parametric MIMO Channel Data Including Path Powers," {\it{IEEE Vehicular Technology Conference, Montreal, QC, Canada}}, 2006, pp. 1-5.

\bibitem{Ref46}
R. He et al., ``A Kernel-Power-Density-Based Algorithm for Channel Multipath Components Clustering," {\it{IEEE Transactions on Wireless Communications}}, vol. 16, no. 11, pp. 7138-7151, 2017.

\bibitem{Ref47}
U. Maulik and S. Bandyopadhyay, ``Performance evaluation of some clustering algorithms and validity indices," {\it{IEEE Transactions on Pattern Analysis and Machine Intelligence}}, vol. 24, no. 12, pp. 1650-1654, 2002.

\bibitem{Ref48}
N. Czink, P. Cera, J. Salo, E. Bonek, J. -p. Nuutinen and J. Ylitalo, ``A Framework for Automatic Clustering of Parametric MIMO Channel Data Including Path Powers," {\it{IEEE Vehicular Technology Conference, Montreal, QC, Canada}}, 2006, pp. 1-5.

\bibitem{Ref49}
E. Schubert, M. Kunert, W. Menzel, J. Fortuny-Guasch and J. -M. Chareau, ``Human RCS measurements and dummy requirements for the assessment of radar based active pedestrian safety systems," {\it{2013 14th International Radar Symposium (IRS), Dresden, Germany}}, 2013, pp. 752-757.
\bibitem{Ref50}
T. Motomura, K. Uchiyama and A. Kajiwara, ``Measurement results of vehicular RCS characteristics for 79GHz millimeter band," {\it{2018 IEEE Topical Conference on Wireless Sensors and Sensor Networks (WiSNet), Anaheim, CA, USA}}, 2018, pp. 103-106.

\bibitem{Ref51}
I. Matsunami, R. Nakamura and A. Kajiwara, ``RCS measurements for vehicles and pedestrian at 26 and 79GHz," {\it{2012 6th International Conference on Signal Processing and Communication Systems, Gold Coast, QLD, Australia}}, 2012, pp. 1-4.





\end{thebibliography}
\end{document}